\pdfoutput=1

\documentclass[journal]{IEEEtran}
\ifCLASSINFOpdf
\else
\fi
\usepackage{amsmath}
\usepackage{graphicx}
\usepackage{subfigure}
\usepackage{amsmath}
\usepackage{algorithm}
\usepackage{algpseudocode}
\usepackage{cite}
\usepackage{multicol}
\usepackage{multirow}
\usepackage{mathtools}
\usepackage{amssymb}
\usepackage{color}
\usepackage{bm}
\usepackage{indentfirst}
\usepackage{url}
\usepackage{nomencl}
\usepackage{booktabs}
\usepackage{cite}

\begin{document}
%
\title{HVAC Energy Cost Optimization for a Multi-zone Building via A Decentralized Approach}
%
%
%

\author{Yu~Yang,~\IEEEmembership{Student Member,~IEEE,}
	Guoqiang~Hu,~\IEEEmembership{Senior Member,~IEEE,}
	and~Costas~J.~Spanos,~\IEEEmembership{Fellow,~IEEE}
	\thanks{This  work  was  supported  by  the  Republic  of  Singapore’s  National  Research  Foundation  through  a  grant  to  the  Berkeley  Education  Alliance  for  Research  in  Singapore
		(BEARS)  for  the  Singapore-Berkeley  Building  Efficiency  and  Sustainability  in  the
		Tropics  (SinBerBEST)  Program.  BEARS  has  been  established  by  the  University  of  California,  Berkeley  as  a  center  for  intellectual  excellence  in  research  and  education  in
		Singapore.}
	\thanks{Yu Yang is with SinBerBEST, Berkeley Education 	Alliance for Research in Singapore, Singapore 138602 e-mail: (yu.yang@bears-berkeley.sg).}
	\thanks{Guoqiang Hu is with the School 	of Electrical and Electronic Engineering, Nanyang Technological University,
		Singapore, 639798 e-mail: (gqhu@ntu.edu.sg).}
	\thanks{Costas J. Spanos is with the Department of Electrical Engineering and 	Computer Sciences, University of California, Berkeley, CA, 94720 USA email: (spanos@berkeley.edu).}
}
\maketitle

\begin{abstract}
	
	It has been well acknowledged that buildings account for a large proportion of the world’s energy consumption. However, the energy use of buildings, especially the heating, ventilation and air-conditioning (HVAC), is far from being efficient. There still exists a dramatic potential to save energy through improving building energy efficiency. Therefore, this paper studies the control of HVAC system for multi-zone buildings with the objective to reduce energy consumption cost  while satisfying thermal comfort. In particular, the thermal couplings due to the heat transfer between the adjacent zones are incorporated in the optimization. Considering that a centralized method is generally computationally prohibitive for large buildings, an efficient decentralized approach is developed, based on the Accelerated Distributed Augmented Lagrangian (ADAL) method \cite{chatzipanagiotis2017convergence}. To evaluate the performance of the proposed method, we first compare it with a centralized method, in which the optimal solution of a small-scale problem can be obtained. We find that this decentralized approach can almost approach the optimal solution of the problem. Further, this decentralized approach is compared with the Distributed Token-Based Scheduling Strategy (DTBSS) \cite{radhakrishnan2016token}. The numeric results reveal that when the number of zones is relatively small (less than 20), the two decentralized methods can achieve a comparable performance regarding the cost of the HVAC system. However, with an increase of the number of zones in the building, the proposed decentralized approach demonstrates better performance with a considerable reduction of the total cost. Moreover, the decentralized approach proposed in this paper demonstrate better scalability with less average computation required.
\end{abstract}

\renewcommand\abstractname{Note to Practitioners}
\begin{abstract} 
Buildings accounts for a large proportion of the world’s energy consumption, especially the  HVAC system. How to improve the energy efficiency of HVAC system  has been recognized as an important and urgent problem for {{a}} sustainable  future. Motivated by this important problem, this paper is focused on the  intelligent  control  of HVAC system for multi-zone buildings with the objective to reduce energy consumption  cost  for  the HVAC system while maintaining  zone  thermal comfort. Considering that a centralized method usually encounters computation difficulties with a large number of zones, this paper {{aims to}} develop an efficient decentralized {{solution}}.   In terms of the problem,  there usually exist various couplings between different zones, which  both arise from the heat transfer between the neighbouring zones and the operation limits of the HVAC system.   Moreover,  this problem is  nonconvex and nonlinear.   Therefore, it is generally  difficult to find an existing decentralized or distributed method, which are mostly established for convex optimization problems.  Motivated by the recent progress on decentralized or distributed optimization for nonconvex problems, this paper proposes an efficient decentralized approach, which mainly contains three steps.  In the first step, the original optimization problem is relaxed by introducing some auxiliary decision variables.   In the relaxed optimization problem, there only exist linear constraints, however, the recursive feasibility of the solution can't be guaranteed. In the second step,  the ADAL method \cite{chatzipanagiotis2017convergence}  is applied to solve the relaxed optimization problem in a decentralized manner. Specifically,  each individual zone can determine its local decision variables by solving a small-scale subproblem in parallel at each iteration.  In the last step, a  heuristic methods is  proposed  to recover the recursive feasibility of the solution for the original optimization problem by exploring the structures of the problem. In this paper, the performance of this decentralized approach is first evaluated through comparison with a centralized method, in which the optimal solution can be obtained.  We find that this decentralized approach can almost approach the optimal solution of the problem.  Moreover, we compare the decentralized approach with  the  DTBSS method   \cite{radhakrishnan2016token}. The numeric results reveal  that the decentralized approach  proposed {{in}} this paper   demonstrates better performance both in reducing the cost of the HVAC systems as well as in improving the computational efficiency {{when the number of zones is large}}. 
\end{abstract}


%
\IEEEpeerreviewmaketitle

\section{Introduction}
%
%
%
%
\IEEEPARstart{I}{t}  has been well acknowledged that  buildings are responsible for a large proportion of the world's  energy consumption. 
Specifically,  {{about  $40\%$ of the primal energy and   $70\%$ of the electricity}}  is consumed in buildings  \cite{ku2015automatic}. 
In particular, about $40\%$-$50\%$ of {{this consumption}}  is attributed to heating, ventilation and air-conditioning (HVAC) systems  \cite{ku2015automatic}.  In tropical countries, like Singapore, Malaysia, and India etc.,  the proportion of  energy consumption caused by HVAC is even higher  \cite{radhakrishnan2016token}.   However,  {{the energy consumption in buildings}}, especially the HVAC {{systems}},  is far from  {{being efficient}}. There still exists  dramatic  potential to reduce  unnecessary energy consumption through improving  building energy efficiency \cite{radhakrishnan2016token}. 

Over the years, the  control of HVAC systems {{and}}  the management of building energy systems have been extensively {{investigated via}} various  methods including {{model predictive control}} (MPC) \cite{ma2012fast, ma2012model, ma2015stochastic, afram2014theory},  mixed-integer linear programming (MILP) \cite{xu2017pmv, xu2015supply},  sequential quadratic programming (SQP) \cite{kelman2011bilinear, sun2005optimal,  maasoumy2012total},  intelligent control based on fuzzy logic or genetic algorithm \cite{killian2016cooperative, nassif2005optimization, wang2000model}, and  rule-based methods \cite{mitsios2009developing, yoon2014dynamic}.  However, most of these methods are carried out in a centralized manner.  
For  large buildings with multiple zones,  {{these methods can be computationally prohibitive or}} not scalable. 
Therefore,  to deal with such situation, this paper focuses on developing an efficient  decentralized approach for  thermal comfort control of HVAC systems  for large multi-zone buildings. 

{{It's challenging}} to achieve optimal control of {{an}}  HVAC system.
For multi-zone buildings,  where a number of thermal zones  share a central HVAC system, the problem is  more complicated  and  challenging. 
\emph{First}, there  exist various decision variables.  Both the temperature and  mass flow rates for all zones  {{need}} to be coordinated to save energy while {{satisfying}} the indoor thermal comfort requirements.
\emph{Second},  there  exist various couplings between different thermal zones.  The couplings both arise from the heat transfer between the adjacent zones and the operation limits of the HVAC system. 
Therefore, there usually exist various temporally and spatially coupled constraints. 
\emph{Third}, the problem is nonlinear and nonconvex.  {{More specifically}}, both the system dynamics  and the global objective function of the problem  are  nonlinear and nonconvex.   
Most of the existing decentralized or distributed  methods, such as  distributed primal-dual subgradient methods \cite{yuan2016regularized}, dual decomposition \cite{terelius2011decentralized},  distributed Alternating Direction Method of Multipliers (ADMM) \cite{boyd2011distributed}  can't be directly applied to solve the problem. Because these decentralized or distributed methods are generally established for convex problems and some of them even can only accommodate  linear constraints.  

{{This paper  investigates  the  control  of HVAC systems  in multi-zone buildings}}   with the objective to reduce energy cost  while guaranteeing  indoor thermal comfort, which incorporates the thermal couplings due to  heat transfer between adjacent zones. 
The main contributions of the paper are  summarized as follows. 
{{Based}}  on decentralized optimization for nonconvex problems with some special structures \cite{chatzipanagiotis2017convergence, magnusson2016convergence},  
an efficient decentralized approach for the control of multi-zone  HVAC systems   is developed  in this paper.  
This decentralized approach mainly contains three steps. 
Specifically, the first step is to relax the original optimization problem. This can be achieved by defining some  auxiliary decision variables.  In the relaxed optimization problem, there only exist linear constraints, however, the recursive feasibility of the solution can't be guaranteed.  In the second step, an Accelerated Distributed Augmented Lagrangian (ADAL) method proposed in  \cite{chatzipanagiotis2017convergence} is applied to solve the nonconvex relaxed optimization problem with only linear constraints. 
Last but very important, the recursive feasibility of the solution is recovered  from  the optimal solution of the relaxed optimization problem through an  heuristic method.  
In this decentralized approach,  all  thermal  zones  can determine their  own  mass flow rates and zone temperature by solving  local subproblems in parallel at each iteration,  therefore this  method is scalable to large buildings {{with a large number of zones}}. 
To evaluate the performance of {{this}} decentralized approach, we first compare it with {{a}}  centralized method,   in which the optimal solution can be {{obtained.}}  We find that the decentralized approach can almost approach the optimal solution of the problem.  
Moreover, this decentralized approach is compared with a distributed Token-Based Scheduling Strategy (TBSS) proposed in \cite{radhakrishnan2016token}.  The numeric results demonstrate that  when the number of zones is relatively small (less than $20$), the decentralized approach can achieve a comparable performance compared with the distributed TBSS method with  about $2\%\sim 4\%$ reduction of the cost. 
However, the decentralized approach outperforms the distributed TBSS  both in  reducing the  cost of the HVAC system  and in reducing  the average computation with a large number of zones in the building.



The remainder of this paper is outlined {{as follows}}. In Section II, the related works  regarding the decentralized or distributed control of HVAC systems  for multi-zone buildings are reviewed. In Section III, the problem formulation is presented.   In Section IV, the decentralized approach is introduced. In Section V, the performance  and scalability of the decentralized approach is validated through case studies.  In Section VI, we briefly conclude this paper.

\section{Related Works}\
As aforementioned, there already exist various  centralized  methods for   {{control of}}  HVAC systems.
{{However, decentralized or distributed methods  for the control of HVAC system in multi-zone buildings have not been well studied.}}
Generally,  the existing  decentralized or distributed  methods   can be divided into  {{two categories}} based on the decision variables. The first category is mainly focused on zone temperature regulation \cite{morocsan2010building, gupta2017incentive, morosan2011distributed}.  
In this category of works,   the zone temperature are regarded as the decision variables. However,  {{how to control the HVAC systems}}  to achieve the desired zone temperature is circumvented and not discussed. 
In these studies,  the zone temperature dynamics are usually assumed to be linear and the problems  are convex. 
Decentralized methods  based on decoupling \cite{morocsan2010building},  incentive price \cite{gupta2017incentive}  and  Dantzig-Wolf decomposition \cite{morosan2011distributed} have been applied to solve the problems. 
In the other category, the control of the HVAC system is directly discussed.  In this case, the thermal  dynamic of each individual zone  is  usually nonlinear, which not only depends on its own local decision variables but also {{depends on those of}}  the adjacent zones due to thermal couplings. 
The problem is usually very complicated and difficult. 
 {{In the literature,}}  most of the existing works ignore the thermal couplings  between the adjacent zones \cite{yu2018distributed, hao2017transactive}.  
{{ In \cite{yu2018distributed},  a real-time decentralized  algorithm based on Lyapunov optimization and some approximation techniques was proposed. }} 
 {{In \cite{hao2017transactive}, to}} protect privacy, a decentralized method based on a  multi-level virtual market was proposed. 
Different from the above works where the heat transfer between neighbouring thermal zones is ignored,  
 {{\cite{radhakrishnan2016token, radhakrishnan2017learning}}} regarded the heat transfer from the adjacent zone as external thermal disturbances  that {{are}} measured through sensors or learned from data at the beginning of each planning horizon. 
To cope with the difficulties and challenges to solve the problem, 
a hierarchical distributed method was proposed, in which the original optimization problem was divided into three-level subproblems  and each corresponds to a part of the global objective function.
Different from those mentioned above,  \cite{zhang2016decentralized, zhang2017decentralized} studied the steady-state  temperature regulation of  the  HVAC  system for an energy-efficient building via a decentralized primal-dual gradient method while relaxing the global objective function.  The problem was formulated as a static optimization problem without considering the thermal dynamics and the thermal coupling among the neighboring zones.

Generally speaking,  in most of the related works, the thermal couplings  among the neighbouring or adjacent zones are ignored or not explicitly considered, which may lead to the  degradation of  performance  in practice. 
On one hand,  the actual energy cost of {{HVAC}} may rise  due to the  heat transfer between the neighbouring zones ignored in the optimization. 
On the other hand, the realistic thermal condition of each individual zone may deviate from the comfortable range. 
{{An attempt was reported in \cite{wang2017distributed}, which}}  studied the distributed MPC strategy for HVAC system in a multi-zone building, where the thermal couplings between the neighbouring zones are explicitly discussed.
To cope with the difficulties due to the nonlinearity and nonconvexity, a distributed  ADMM method was applied  based on some convexity approximation. 

 
Complementary to the existing works, {{this paper considers  the thermal couplings  among the neighbouring zones.}}  To improve scalability of the method, an efficient decentralized approach based on ADAL is developed.

\section{Problem Formulation}

 \subsection{HVAC  Systems  for Multi-zone Buildings}
 
\begin{figure}
	\centering
	\includegraphics[width=3.4 in]{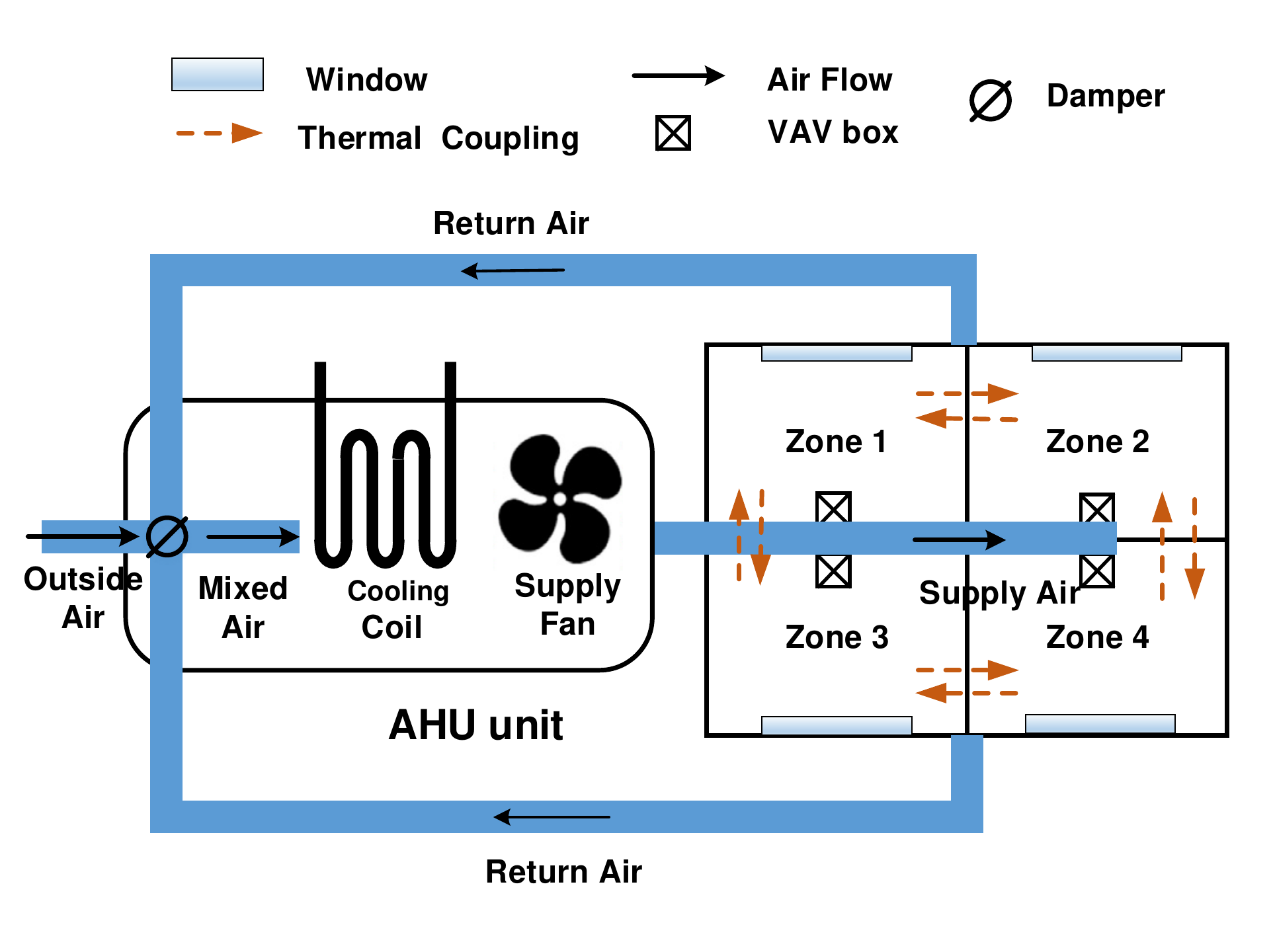}\\
	\caption{The schematic of  an HVAC system in a  multi-zone building.}\label{system architecture}
\end{figure}

 A typical schematic of {{an  HVAC system}}  in a multi-zone building is shown in Fig. \ref{system architecture}. 
 The main parts of the system include the Air Handling Unit (AHU), the Variable Air Volume (VAV) box, and the 
 chiller. The central AHU is shared by multi-zones, which is equipped with a damper, a cooling/heating coil and a supply fan. The damper is responsible for mixing the return air from inside and the freash air from outside. 
 The heating/cooling coil can cool down/heat up the mixed air to a setpoint temperature. 
 Without loss of generality, this paper considers the cooling mode of the HVAC system. 
Generally, {{the temperature of the supply air out of AHU is}}  $12$-$16^\circ$C in the cooling mode. 
 Besides, there is a local VAV box related to each zone, which  
 consists of a damper and a heating coil. This damper  
 can regulate  {{air flow rate supplied}} to the zone, while the heating coil can reheat the supply air before supplied to the zone if necessary (in this paper, the heating coil of the local VAV box is not discussed). More details regarding HVAC systems  for multi-zone buildings  can  refer to  \cite{kelman2011bilinear, zhang2017decentralized}. 
 
{{In terms of the control of HVAC systems  in multi-zone buildings,  one important problem is how to improve energy efficiency while 
				satisfying the  thermal comfort  of each zone.    This problem is complicated and difficult due to the various couplings between different zones. }}
On one hand, the thermal dynamics of different zones are  usually coupled with each other due to heat transfer. 
On the other hand,  {{the control of the HVAC system shared by different zones  is generally coupled by the operation limits and the nonlinear energy consumption cost of the system.}} 
  
In this paper, the problem is discussed in a {{discrete-time}} framework. The optimization horizon (one day) is equally divided into $T=48$ stages, each corresponding to a decision interval of $\Delta_t=30$ mins.  

\subsection{Zone Thermal  Dynamics}
In this paper, we consider a multi-zone building with $I$ thermal zones, which are indexed by $\mathcal{I}=\{1, 2, \cdots, I\}$. The control of the  HVAC system over the optimization horizon $\mathcal{T}=\{0, 1, \cdots, T-1\}$ is studied. 
The thermal dynamics of zone $i$ ($\forall i \in \mathcal{I}$) can be described based on the  Resistance-Capacitance (RC) network  \cite{lin2012issues, maasoumy2011model}, i.e.,
\begin{equation} \label{zone temperature dynamics}
\begin{split}
C_i  (T^i_{t+1}&-T^i_t)=\sum_{j\in \mathcal{N}_i} \frac{T^j_t-T^i_t}{R_{ij}}  \Delta_t\\
&+\frac{T^o_t-T^i_t}{R_{oi}} \Delta_t+c_p m^{zi}_t 
(T^{c}_t-T^i_t)\Delta_t+Q^i_{t} \Delta_t
\end{split}
\end{equation}
where $C_i$ is the heat capacity of the indoor air in zone $i$. $T^i_t$ is the temperature of zone $i$ at time $t$.
$T^o_t$ denotes the outside air temperature at time $t$. $R_{oi}$ is the thermal resistance of the window between zone $i$ and the outside.  $R_{ij}$ denotes the thermal resistance between zone $i$ and zone $j$. $\mathcal{N}_i$ denotes the collection of adjacent zones of zone $i$. 
$c_p$ is the specific heat of air. $m^{zi}_t$ denotes the air flow rate {{supplied}} to zone $i$.
$Q^i_t$ denotes  the  total heat generation in  zone $i$, which may result from  occupants,  devices and solar radiation, etc.

If we define $A^{ii}=1-(\sum_{j\in \mathcal{N}_i} \frac{\Delta_t}{R_{ij}C_i}+\frac{\Delta_t}{C_i R_{oi}})$, $A^{ij}=\frac{\Delta_t}{C_i R_{ij}}$,
$C^{ii}=-\frac{\Delta_t \cdot  c_p}{C_i} $, and $D^{ii}_t=\frac{\Delta_t T^o_t}{C_i R_{oi}}+\frac{\Delta_t \cdot Q^i_t}{C_i}$. the zone thermal dynamics in (\ref{zone temperature dynamics}) can be equivalently described as 
\begin{equation} 
T^i_{t+1}=A^{ii}T^i_t+\sum_{j\in \mathcal{N}_i} A^{ij} T^j_t+C^{ii} m^{zi}_t (T^i_t-T^c_t)+D^{ii}_t
\end{equation}




\subsection{The AHU}
As introduced in Section II-A, the AHU  is responsible for cooling  down  the mixed air to a setpoint temperature before supplied to each individual zone. 
The main parameters related to the AHU include 
1) {{$d_r$}}  ($0\!\leq\! d_r \!\leq \!1$): the fraction of {{the return air from inside.}}
2) $T^r_t$: the average temperature of the return air from inside.
3) $T^m_t$: the  average temperature of the mixed air before supplied to the AHU.
4) $T^c_t$: the setpoint temperature of the suppy  air by  the AHU. 
The settings  of $d_r$ and $T^c_t$ are usually determined based on experience. 
The  average temperature of  the return air from inside at time $t$  is determined by 
\begin{equation}
\begin{split}
T^r_t=\frac{\sum_{i=1}^I m^{zi}_t T^{i}_t}{\sum_{i=1}^I m^{zi}_t}
\end{split}
\end{equation}
And  the average  temperature of the mixed air that supplied to the AHU at time $t$ can be determined by
\begin{equation} \label{return air}
\begin{split}
T^m_t&=(1-d_r)T^o_t+d_r T^r_t\\
&=(1-d_r)T^o_t+d_r \frac{\sum_{i=1}^I m^{zi}_t T^{i}_t}{\sum_{i=1}^I m^{zi}_t}
\end{split}
\end{equation}


As aforementioned, the AHU  is mainly composed {{of}} the cooling coil and the supply fan. 
Therefore, the  main energy consumption of the AHU consists of two parts. The first part is caused by the cooling coil, which can be determined by 
\begin{equation} \label{AHU cooling power}
\begin{split}
  P^c_t= c_p \eta (\sum_{i=1}^I m^{zi}_t) (T^m_t-T^c_t)
\end{split}
\end{equation}\
where $\eta$ is  the  reciprocal  of the coefficient  of  performance  (COP)  of  the  chiller, which captures the ratio of provided heating
or cooling to the total consumed electrical energy.

By substituting (\ref{return air}) into (\ref{AHU cooling power}), we have 
\begin{equation} \label{AHU cooling power2}
\begin{split}
P^c_t\!\!=\!\!c_p \eta (1\!-\!d_r)\!\!\sum_{i=1}^I \!m^{zi}_t\!(T^o_t\!-\!T^c_t)\!+\!c_p \eta d_r \sum_{i=1}^I\! m^{zi}_t (T^i_t\!-\!T^c_t)\\
\end{split}
\end{equation}

From (\ref{AHU cooling power2}), we note that the cooling power of the AHU can be {{divided into}}  two parts. The first part is to  cool down the proportion of  the outside fresh air to the setpoint temperature. And the second part  is to cool down the proportion of  the return air from inside. 

{{The second part of the energy consumption results from the supply fan. }} According to \cite{hao2017transactive} and \cite{wang2017distributed}, the energy consumption of the supply fan depends on the cube of the total zone  air flow rate, i.e., 
\begin{equation} \label{fan power}
\begin{split}
P^f_t=\kappa_f (\sum_{i=1}^I m^{zi}_t)^3
\end{split}
\end{equation}
where $\kappa_f$ is a fixed parameter that captures the efficiency of the supply fan. 

Therefore, the total power consumed by  the HVAC system  at time $t$ can be calculated as
\begin{equation} \label{stage cost}
\begin{split}
	&P^c_t+P^f_t\!=\!c_p (1-d_r)\sum_{i=1}^I m^{zi}_t (T^o_t-T^c_t)\!\\
	&\quad \quad \quad +\!c_p d_r\sum_{i=1}^I m^{zi}_t (T^i_t-T^c_t) +\kappa_f (\sum_{i=1}^I {m^{zi}_t})^3\\
\end{split}
\end{equation}

From (\ref{stage cost}), we  note that  {{the power consumption of }} the HVAC system is a nonlinear function with respect to the decision variables $m^{zi}_t$ and state variables $T^i_t$  of  each individual  zone $i$ ($i\in\mathcal{I}$).

\subsection{System Constraints}

In terms of the control of {{the HVAC system}}  in  multi-zone building, there exist various constraints. 
First of all, the thermal comfort requirement of each individual zone should be guaranteed. In this paper, we use temperature ranges  to describe the thermal comfort requirements of different zones, thus we have
\begin{equation} \label{temperature range}
\begin{split}
	\underline{T}^i & \leq T^i_t \leq \overline{T}^i, \quad \forall i \in \mathcal{I}, ~t \in \mathcal{T}
	\end{split}
\end{equation}
where $\underline{T}^i$ and $\overline{T}^i$ represent the lower and upper bound of the comfortable temperature  for zone $i$, respectively. 
To accommodate personalized thermal comfort, $\underline{T}^i$ and $\overline{T}^i$ may be different for different zones.

Also, the zone air flow rate  is usually bounded, i.e.,
\begin{equation} \label{each zone air flow rate bound}
\begin{split}
&\underline{m}^{zi} \leq m^{zi}_t \leq \overline{m}^{zi}, \quad \forall i \in \mathcal{I}, ~t \in \mathcal{T}\\
\end{split}
\end{equation}
where $\underline{m}^{zi}$ and $\overline{m}^{zi}$ {{denote}}  the lower and upper bound of the air flow rate for zone $i$, respectively. The lower and upper bound of the zone air flow rates are generally determined by the pressure supplied by the fan in the duct system \cite{radhakrishnan2016token}.

Besides, there usually exist operation limits for the central AHU that shared by the multi-zones, i.e.,  
\begin{equation} \label{total air flow rate bound}
\begin{split}
&\sum_{i=1}^I m^{zi}_t \leq  \overline{m}, ~\forall t \in \mathcal{T}\\
\end{split}
\end{equation}
where $\overline{m}$ denotes the {{maximum total}} air flow rate that can be supplied  by the AHU at the same time. 

\subsection{The Optimization Problem}
{{To improve energy efficiency, the problem is how to reduce the energy cost of the HVAC system while guaranteeing the thermal comfort  of different zones. Therefore, the optimization problem can be described as follows.  }}
\begin{align}
 \label{obj} \min_{m^{zi}_t, T^i_t} J=\sum_{t=0}^{T-1}  c_t \cdot \Big\{  P^c_t +P^f_t
 \Big\} \cdot \Delta_t  \tag{\textbf{P1}}\\
 \textrm{\textbf{{Constraints:}}}~ (\ref{zone temperature dynamics}), (\ref{temperature range})-(\ref{total air flow rate bound}) \quad \quad  \notag
\end{align}
where $c_t$ denotes  the electricity price at time $t$. 

In (\ref{obj}),  \label{key}the total energy consumption  cost of the HVAC system over the optimization horizon $\mathcal{T}$ is selected as the global objective function.  
{{We note  that   this  global objective function is non-separable and  nondecomposable with respect to the thermal zones. }}
Becides,  there exist various coupled nonlinear constraints due to the thermal coupling among different zones and the operation limits of the HVAC system. 
{{The  problem (\ref{obj}) is a nonlinear and nonconvex optimization problem, }}  which is difficult  to find an existing  decentralized or distributed  method to solve it. 
Because most of the existing decentralized or distributed  methods \cite{duchi2012dual, nedic2009distributed, zhu2012distributed, yuan2016regularized, terelius2011decentralized, boyd2011distributed}, such as  distributed primal-dual subgradient methods \cite{yuan2016regularized}, dual decomposition \cite{terelius2011decentralized},  distributed ADMM \cite{boyd2011distributed} are developed for convex optimization problems and some of them  can only tackle  linear constraints.

\section{Decentralized Approach}
As aforementioned, it is challenging and nontrivial to find  an existing  decentralized or distributed  solution method for  (\ref{obj}) due to the nonconvexity and nonlinearity of the problem. 
To cope with the difficulties, 
this paper  proposes an efficient decentralized approach to solve  problem (\ref{obj}).  
Generally, this decentralized approach mainly contains three steps. 
In the first step, the original optimization problem (\ref{obj}) is relaxed by introducing some  appropriate auxiliary decision variables. {{For  the relaxed optimization problem,}}   there only exist linear (coupled and decoupled) constraints. However,  the recursive feasibility of the solution  {{cannot}}  be guaranteed.  
In the second step, an  ADAL method \cite{chatzipanagiotis2017convergence} is applied to solve the nonconvex  relaxed optimization problem in a decentralized manner.  Last but very important, the third step is  focused  on recovering  the recursive feasibility of the solution  for (\ref{obj}).

 {{To deal with the nonlinear constraints of problem (\ref{obj}), }}  we first introduce  the following auxilliary 
decision variables,  i.e., $X^i_t=m^{zi}_t (T^i_t-T^c_t)\geq 0$  ($\forall i \in \mathcal{I}, t \in \mathcal{T}$) and  $Y_t=\sum_{i=1}^Im^{zi}_t$ ($ \forall t \in \mathcal{T}$).  
We note that  $X^i_t$ can be regarded as the ``cooling power" {{supplied to}} zone $i$ at time $t$,  
and $Y_t$ can be regarded as the {{total air flow rate supplied by the AHU to all zones in the building at time $t$.}}

According to \cite{mccormick1976computability} and \cite{wang2017distributed}, these  auxiliary decision variables $X^i_t$ ($\forall i\in\mathcal{I}$, $t\in \mathcal{T}$) are  bounded by their  convex and concave envelopes, respectively. Therefore, we have 
\begin{equation} \label{(12)}
\begin{split}
&	X^i_t=m^{zi}_t (T^i_t-T^c_t) \\
	& \!\geq\!\max\!\Big\{ \underline{m}^{zi}\!(T^i_t\!-\!T^c_t)\! +\!m^{zi}_t (\underline{T}^i\!-\!T^c_t)\!-\!\underline{m}^{zi}(\underline{T}^i-T^c_t), \\
	&\quad  \quad \quad   \overline{m}^{zi} (T^i_t\!-\!T^c_t)\!+\!m^{zi}_t(\overline{T}^i\!-\!T^c_t)\!-\!\overline{m}^{zi}(\overline{T}^i\!-\!T^c_t)\Big\},\\
&	X^i_t=m^{zi}_t (T^i_t-T^c_t) \\
& \leq \min \Big\{  m^{zi}_t(\overline{T}^i\!-\!T^c_t)\!+\!\underline{m}^{zi}(T^i_t\!-\!T^c_t)\!-\!\underline{m}^{zi}(\overline{T}^i\!-\!T^c_t), \\
&\quad  \quad \quad   \overline{m}^{zi}(T^i_t\!-\!T^c_t)\!+\!m^{zi}(\underline{T}^i\!-\!T^c_t)\!-\!\overline{m}^{zi} (\underline{T}^i-T^c_t)\Big\}, \\
&\quad \quad \quad \quad \forall i \in \mathcal{I},~ t \in \mathcal{T}.\\
\end{split}
\end{equation}

Therefore, by introducing these auxiliary decision variables,  {{we can obtain the following relaxed optimization problem (\ref{P2}) regarding the original optimization  
		problem (\ref{obj}):}} 
\begin{align}
&\min_{  m^{zi}_t, T^i_t,  X^i_t, Y_t}  J=\sum_{t=0}^{T-1} c_t \cdot \Big\{c_p \eta (1-d_r) (T^o_t-T^c_t) Y_t  \quad \quad \quad \notag\\
&\label{P2}\quad \quad \quad \quad \quad  \quad +\kappa_f (Y_t)^3+c_p \eta d_r \sum_{i=1}^I X^i_t \Big\}\cdot \Delta_t  \quad  \tag{\textbf{P2}}\\ \notag
& \textrm{\textbf{{subject to}}}  \notag\\
&\label{14a} T^i_{t+1}=A^{ii}T^i_t+\sum_{j\in \mathcal{N}_i} A^{ij} T^j_t+C_{ii}X^i_t +D^{ii}_t, \tag{13a} \\
&\label{14b} \underline{m}^{zi} \leq m^{zi}_t \leq \overline{m}^{zi}, \tag{13b} \\ \notag
\end{align}
\begin{align}
&\label{14c}  \underline{T}^{i} \leq T^{i}_t \leq \overline{T}^{i},  \tag{13c}\\
&\label{14d} X^i_t \!\geq\! \underline{m}^{zi}(\!T^i_t\!-\!T^c_t)\! +\!m^{zi}_t (\underline{T}^i-T^c_t)-\underline{m}^{zi}(\underline{T}^i-T^c_t), \tag{13d}\\
&\label{14e}   X^i_t \!\geq \! \overline{m}^{zi} (T^i_t-T^c_t)\!+\!m^{zi}_t(\overline{T}^i\!-\!T^c_t)\!-\!\overline{m}^{zi}(\overline{T}^i-T^c_t), \tag{13e}\\
&\label{14f} X^i_t\! \leq \!m^{zi}_t(\overline{T}^i\!-\!T^c_t)\!+\!\underline{m}^{zi}(T^i_t-T^c_t)-\underline{m}^{zi}(\overline{T}^i-T^c_t), \tag{13f}\\
&\label{14g}   X^i_t \!\leq \! \overline{m}^{zi}(T^i_t-T^c_t)\!+\!m^{zi}(\underline{T}^i\!-\!T^c_t)\!-\!\overline{m}^{zi} (\underline{T}^i-T^c_t), \tag{13g}\\
&\quad \quad \quad \quad \quad \forall i \in \mathcal{I}, ~ t \in \mathcal{T}.  \notag\\
&\label{14h}\sum_{i=1}^I m^{zi}_t \leq Y_t,  \tag{13h} \\
& \label{14i} \sum_{i=1}^I m^{zi}_t \leq  \overline{m}, \forall t \in \mathcal{T}.\tag{13i}\\  
\end{align}

In (\ref{P2}), Constraints (\ref{14d})-(\ref{14g})  correspond to 
Constraints (\ref{(12)}), which relate to the auxiliary decision variable $X^i_t$. And  Constraints
(\ref{14h}) relate  to the auxiliary decision variable $Y_t$.  {{We note that problem (\ref{P2}) is  still nonconvex due to the global objective function,  however, there only exist linear constraints. }}
This problem can be tackled by the ADAL method \cite{chatzipanagiotis2017convergence}, which has been established for 
problems with nonconvex global objective function but with only linear coupled constraints. 
Therefore, in the second step, we apply the ADAL method to solve  problem (\ref{P2}) in a decentralized manner.  
{{More specifically}},  we assume that  there are $I+1$ agents, which are indexed by $\mathcal{I} \cup \{0\}$. 
The collection of the agents  $\mathcal{I}$ corresponds  to the $I$  zones in the building. 
In addition, we define a virtual  Agent  $0$,   to regulate the total  air flow rate supplied by the AHU.

For each Agent $i$  that related to zone $i$ ($\forall i \in \mathcal{I}$), the local decision variables at time $t$ can be  gathered  as  
\begin{equation}
\begin{split}
	\bm{x}^i_t=&\Big(T^i_t,  m^{zi}_t,  X^i_t\Big)^T,~\forall i \in \mathcal{I}, t \in \mathcal{T}. 
\end{split}
\end{equation}
 
For  Agent $0$, the local decision variable at time $t$ is defined as

 \begin{equation}
 	\bm{x}^0_t=Y_t, ~\forall t \in \mathcal{T}.
 \end{equation}

For notation, we use the vectors  $\bm{x}^i=[(\bm{x}^i_0)^T, (\bm{x}^i_1)^T, \cdots, $ $(\bm{x}^i_{T-1})^T ]^T$ to represent the local decision variables  of Agent $i$ ($\forall i \in \mathcal{I} \cup\{0\}$)  over the optimization horizon $\mathcal{T}$.

Accordingly, the global objective function in (\ref{P2}) can be decomposed with respect to the agents. 
{{
\begin{eqnarray}
\begin{split}
&\textrm{For $i \in \mathcal{I}$,  we have }\\
&\quad \quad J_i( \bm{x}^i)\!\!=\!\sum_{t=0}^{T-1} \!c_t\! \cdot \Big\{ c_p \eta d_r\! X^i_t\! \Big\} \cdot \Delta_t \\
&\textrm{For  $i=0$,  we have} \\
&J_0(\bm{x}^0)\!=\!\!\!\sum_{t=0}^{T-1}\!\!c_t \!\cdot \!\Big\{ c_p \eta (1\!-\!d_r)(T^o_t-T^c_t) Y_t\!+\!\kappa_f (Y_t)^3 \Big\}\! \cdot \!\Delta_t
\end{split}
\end{eqnarray}
}}

It's straightforward that 
$ J=\sum_{i=0}^I J_i(\bm{x}^i)$, 
where $J_i(\bm{x}^i)$ can be regarded as the local objective function of Agent $i$.

Before we can apply the ADAL method to {{solve problem (\ref{P2})}}, we first need to convert the problem to the specific structures as required by the method. First of all,  problem  (\ref{P2})  can be  transformed to (\ref{P3}) as follows. The details of the transformation can be found in \textbf{Appendix} A. 
\vspace{-4mm}
\begin{align} 
&\min_{\bm{x}^i, i=0, 1,2, \cdots, I} \sum_{i=0}^I J_i(\bm{x}_i)  \notag\\
 s.t.\quad &\bm{A}_d^{ii}\bm{x}^i +\sum_{j\in \mathcal{N}_i} \bm{A}_d^{ij} \bm{x}^j=\bm{b}_d^i, \forall  i \in \mathcal{I}. \notag\\
&\label{P3}  \sum_{i=0}^I \bm{B}_d^i \bm{x}^i \leq \bm{0}. \notag \tag{\textbf{P3}}\\
&\sum_{i=1}^I \bm{B}_d^i \bm{x}^i \leq \bm{c}_d.  \notag\\
&\bm{x}^i \in \mathcal{X}^i, \forall i \in \mathcal{I}.  \notag
\end{align}
where we have $$\bm{A}_d^{ii}= \left(      
\begin{array}{cccccc}   
\overline{A}^{ii} & -\mathbf{I}_1  & \bm{0} & \cdots & \cdots& \cdots\\  
\bm{0} &  \overline{A}^{ii} & -\mathbf{I}_1 & \bm{0} &\cdots & \cdots \\  
\bm{0} & \bm{0} & \overline{A}^{ii} & -\mathbf{I}_1  & \bm{0} & \cdots \\
\end{array}
\right) \in \mathbb{R}^{(T-1) \times 3T}, $$
$\bm{A}_d^{ij}=\left( \begin{array}{ccccc}   
\overline{A}^{ij} & \bm{0}  & \cdots & \cdots & \cdots\\  
\bm{0} &  \overline{A}^{ij} & \bm{0} & \cdots & \cdots\\  
\bm{0} & \bm{0} & \overline{A}^{ij} & \bm{0} & \cdots \\
\end{array}
\right) \in \mathbb{R}^{(T-1) \times  3T}, \quad \quad 
$
and  $\bm{b}_d^i\!=\!(\overline{D}^{ii}_0, \overline{D}^{ii}_1, \cdots, \overline{D}^{ii}_{T-1})^T\! \! \in\! \mathbb{R}^{T-1},$
with $\overline{A}^{ii}\!=\!(A^{ii}~~0$ $~~C^{ii})$, $\overline{A}^{ij}=(A^{ij}~~0~~0)$ $(i, j \in \mathcal{I})$.
$\mathbf{I}_1\!=\!(1~0~0)$.  $\overline{D}^{ii}_t=-D^{ii}_t.$         
 
 \noindent
 $\bm{B}_d^i\!=\!\!\!\left(\begin{array}{ccccc}
\overline{B}^i \!& \!\bm{0} \! &\cdots \!& \cdots  & \cdots\\
\bm{0} \! & \! \overline{B}^i \! & \bm{0}\! & \cdots  & \cdots\\
\bm{0} \! &\! \bm{0} \!& \overline{B}^i \! & \bm{0} & \cdots\\
\end{array} \!\!\right) \in \mathbb{R}^{T \times 3T} $, 
with  $\overline{B}^i\!=\! (0~1~0)$  $(\forall i \in \mathcal{I}).$ 
$\bm{B}_d^0=\left(\begin{array}{ccccc}
\overline{B}^0 & \bm{0} &\bm{0} & \cdots\\
\bm{0} & \overline{B}^0 & \bm{0}  & \cdots  \\
\bm{0}  &\bm{0} & \overline{B}^0  & \cdots \\
\end{array} \right) \in \mathbb{R}^{T \times T}$ with $\overline{B}^0=(-1)$, 
$\bm{c}_d \!\!=\!\!(\overline{m}~~\overline{m}~\cdots~\overline{m})^T \in \mathbb{R}^T.$ 
For notation, we use the set $\mathcal{X}^i$ to represent  the collection of  admissible  control trajectories  for  Agent $i$ ($i \in \mathcal{I}$), which can be constructed 
 by the local constraints (\ref{14b})-(\ref{14g}) related to zone $i$. 

Observe (\ref{P3}), we see that there  not only exist coupled equality constraints but also exist coupled inequality constraints,  the latter of which {{can not be}} tackled efficiently by the ADAL method  \cite{chatzipanagiotis2017convergence}.  
To overcome the difficulty, we introduce two other auxiliary decision variables $\bm{s}^1$ and $\bm{s}^2$ as discussed in \cite{chang2016proximal}.
In this case,  (\ref{P3}) is equivalent to (\ref{P4}) as follows. 
\begin{align}
&\min_{\bm{x}^i, i=0, 1,2, \cdots, I, \bm{s}^1, \bm{s}^2 } \sum_{i=0}^I J_i(\bm{x}_i)  \notag\\
s.t. \quad & \bm{A}_d^{ii}\bm{x}^i +\sum_{j\in \mathcal{N}_i} \bm{A}_d^{ij} \bm{x}^j=\bm{b}_d^i, \quad \forall  i\in \mathcal{I}. \notag\\
& \label{P4} \sum_{i=0}^I \bm{B}_d^i \bm{x}^i +\bm{s}^1=\bm{0}. \tag{\textbf{P4}} \\
&\sum_{i=1}^I \bm{B}_d^i \bm{x}^i-\bm{c}_d +\bm{s}^2 =\bm{0}. \notag\\
&\bm{x}^i \in \mathcal{X}^i, ~\forall i \in \mathcal{I}. \notag\\
&\bm{s}^1,   \bm{s}^2 \geq \bm{0}. \notag
\end{align}
where  $\bm{s}^1=[s^1_0, s^1_1, \cdots, s^1_{T-1}]^T\in \mathbb{R}^T$ and $\bm{s}^2=[s^2_0, s^2_1, $ $ \cdots, s^2_{T-1}]^T \in \mathbb{R}^T$ are auxiliary decision variables.

In (\ref{P4}), there only exist coupled linear equality constraints, which corresponds well to the special structures required by the ADAL method  \cite{chatzipanagiotis2017convergence}. 
{{Therefore, in the second step, the ADAL method is applied to tackle problem (\ref{P4}). More specifically, }}
 we first define the following augmented Lagrangian function to eliminate the coupled equality constraints, i.e., 
\begin{eqnarray} \label{Lagrangian global objective function}
\begin{split}
\mathbb{L}_{\rho}&( \bm{x}^0, \bm{x}^1, \cdots, \bm{x}^I, \bm{s}^1, \bm{s}^2,   \bm{\lambda},  \bm{\gamma}, \bm{\eta})=\sum_{i=0}^I J_i\\
&+\sum_{i=1}^I (\bm{\lambda}^i)^T  \Big(  \bm{A}_d^{ii}\bm{x}^i +\sum_{j\in \mathcal{N}_i} \bm{A}_d^{ij} \bm{x}^j-\bm{b}_d^i\Big) \\
&+\sum_{i=1}^I\frac{\rho}{2} \Big\lvert\Big\lvert \bm{A}_d^{ii}\bm{x}^i +\sum_{j\in \mathcal{N}_i} \bm{A}_d^{ij} \bm{x}^j-\bm{b}_d^i \Big\rvert\Big \rvert^2\\
&\!+\!\bm{ \gamma}^T \Big(\sum_{i=0}^I \bm{B}_d^i \bm{x}^i +\bm{s}^1\Big)\!+\!\frac{\rho}{2} \Big \Vert  \sum_{i=0}^I \bm{B}_d^i \bm{x}^i +\bm{s}^1 \Big \Vert^2\\
&\!+\! \bm{\eta}^T \Big(\sum_{i=1}^I \bm{B}_d^i \bm{x}^i\!-\!\bm{c}_d \!+\!\bm{s}^2\Big)\!+\!\frac{\rho}{2} \Big \Vert \sum_{i=1}^I \bm{B}_d^i \bm{x}^i\!-\!\bm{c}_d \!+\!\bm{s}^2 \Big \Vert^2
\end{split}
\end{eqnarray}
where $\bm{\lambda},  \bm{\gamma}$  and $\bm{\eta}$ are Lagrangian multipliers, we have $\bm{\lambda} = [ (\bm{\lambda}^1)^T, (\bm{\lambda}^2)^T, \cdots, (\bm{\lambda}^I)^T]^T$,   with 
$\bm{\lambda}^i \in \mathbb{R}^{(T-1) \times 3T}$, and  $\bm{\gamma},  \bm{\eta} \in \mathbb{R}^T $. $\rho>0$  is a penalty parameter.

Thus, the primal  problem of  (\ref{P4}) can be described as 
\begin{equation} \label{primal problem}
\begin{split}
\min_{\bm{x}^0, \bm{x}^1, \cdots, \bm{x}^I, \bm{s}^1, \bm{s}^2} &\mathbb{L}_{\rho}( \bm{x}^0, \bm{x}^1, \cdots, \bm{x}^I, \bm{s}^1, \bm{s}^2,  \bm{\lambda}, \bm{\gamma}, \bm{\eta})\\
 s. t. \quad & \bm{x}^i  \in \mathcal{X}^i, \forall i \in \mathcal{I}. \\
& \bm{s}^1    \geq \bm{0}. \\
& \bm{s}^2   \geq  \bm{0}. 
\end{split}
\end{equation}

{{When the ADAL method is applied to solve (\ref{P4}),  the primal  problem  (\ref{primal problem})  can be tackled in a decentralized manner. }}
The details of the algorithm  are displayed in \textbf{Algorithm}
\ref{ADAL}. Hereafter, we use the superscript  $k$ to represent the number of iteration.  In \textbf{Algorithm} \ref{ADAL}, the local  objective functions for each zone $i$ ($i \in \mathcal{I} \cup \{0\}$) at iteration $k$ is  defined as follows.   
\begin{equation}
\begin{split}	 
	 & \mathbb{L}^i_{\rho} (\bm{x}^i, \bm{x}^{-i, k}, \bm{s}^{1, k}, \bm{s}^{2, k},  \bm{\lambda}^k, \bm{\gamma}^k, \bm{\eta}^k)=  
	 J_i(\bm{x}^i)\\
	 &+ \sum_{j\in \mathcal{N}_i \cup\{i\}} ({\bm{\lambda}^k_j})^T \bm{A}_d^{ji}\bm{x}^i \\
	 &+\frac{\rho}{2}\sum_{j\in \mathcal{N}_i \cup\{i\}} \Big \Vert  \bm{A}_d^{ji}\bm{x}^i +\sum_{l\in \mathcal{N}_j} \bm{A}^{jl} \bm{x}^{l,k}-\bm{b}_d^j \Big \Vert^2\\
	 &+(\bm{\gamma}^k)^T \bm{B}_d^i \bm{x}^i+\frac{\rho}{2} \Big\Vert \bm{B}_d^i \bm{x}^i+ \sum_{\mathclap{l=0,  l \neq i}}^I \bm{B}_d^l \bm{x}^{l, k} +\bm{s}^{1, k}  \Big\Vert^2 \\
	 &+(\bm{\eta}^k)^T\bm{B}_d^i \bm{x}^i \!+\!\frac{\rho}{2}\Big\Vert  \bm{B}_d^i \bm{x}^i\!+\! \sum_{\mathclap{l=1, l\neq i}}^I \bm{B}_d^l \bm{x}^l\! -\! \bm{c}_d\! +\!\bm{s}^{2, k}  \Big\Vert^2, \\
	 &\quad \quad \quad ~\forall i \in \mathcal{I}.\\
	  \end{split}
	  \end{equation}
	  
	  \begin{equation}
	  \begin{split}
	  &\mathbb{L}^0_{\rho} (\bm{x}^0, \bm{s}^1, \bm{s}^2, \bm{x}^{-0, k}, \bm{\lambda}^k, \bm{\gamma}^k, \bm{\eta}^k) 
	  =J_0(\bm{x}^0)\\
	  &+\bm{\gamma}^T (\bm{B}_d^0  \bm{x}^0+\bm{s}^1)+\frac{\rho}{2} \Big\Vert \bm{B}_d^0 \bm{x}^0 +\bm{s}^1+ \sum_{i=1}^I \bm{B}_d^i \bm{x}^{i, k}   \Big\Vert^2\\
	  &+\bm{\eta}^T \bm{s}^2+\frac{\rho}{2}\Big\Vert  \sum_{i=1}^I \bm{B}_d^i \bm{x}^{i, k} - \bm{c}_d +\bm{s}^2  \Big\Vert^2\\
	  \end{split}
	  \end{equation}
For notation,  we use $\bm{x}^k=( (\bm{x}^{0, k})^T,  (\bm{x}^{1, k})^T, $ $ \cdots, (\bm{x}^{I, k})^T)^T$ to represent the collection of control trajectories  for  all zones at iteration $k$. 
$\bm{x}^{-i, k}=\big ((\bm{x}^{0, k})^T, \cdots,  (\bm{x}^{i-1, k})^T,  (\bm{x}^{i+1, k})^T,  \cdots, $ $(\bm{x}^{I, k})^T\big)^T$ denotes the collection of {{control trajectories}} for all zones  except zone $i$.
In \textbf{Algorithm} \ref{ADAL},  the residual  error of all the coupled constraints is utilized as the stopping criterion, which is defined as 
\begin{equation} \label{residual}
\begin{split}
&r(\bm{x^k}, \!\bm{s}^{1, k}, \!\bm{s}^{2, k})\!=\!\sum_{i=1}^I \Big\Vert \bm{A}_d^{ii}\bm{x}^{i, k} \!+\!\!\sum_{j\in \mathcal{N}_i} \bm{A}_d^{ij} \bm{x}^{j, k}\!-\!\bm{b}_d^i\Big\Vert_2 \\
&\!+\!\Big\Vert  \sum_{i=0}^I \bm{B}_d^i \bm{x}^{i, k} \!+\!\bm{s}^{1, k}  \Big\Vert_2+\Big\Vert \sum_{i=1}^I \bm{B}_d^i \bm{x}^{i, k}\!-\!\bm{c}_d \!+\!\bm{s}^{2, k} \Big\Vert_2
\end{split}
\end{equation}

In \textbf{Algorithm} \ref{ADAL}, it's not difficult to note  that the subproblems related to the $I$  zones are quadratic programming (QP) problems, which can be {{efficiently}}  tackled by many existing toolboxes, such as CPLEX. Though Subproblem $0$ is a nonlinear optimization problem due to the local objective function,  this subproblem is  a small-scale  problem with only simple local constraints. Therefore, Subproblem $0$ also can be solved  efficiently. 
 {{$\epsilon$ is a constant threshold, which is   determined  by  the suboptimality of solution as required. }}

\begin{algorithm}[h] 
	\caption{Accerlerated Distributed Augmented Lagrangian (ADAL)} \label{ADAL}
	\begin{algorithmic}[1]
		\State \textbf{Initialization:} $ k \leftarrow 0$,  set $\bm{\lambda}^0$, $\bm{\gamma}^0$, $\bm{\eta}^0$, $\bm{x}^{i, 0}$ ($\forall i\in \{0\} \cup \mathcal{I}$) and $s^{1, 0}$, $\bm{s}^{2, 0}$.
		\State \textbf{Iteration:}
		\State For $i=0$, obtain $\bm{x}^{0, k+1}$, $\bm{s}^{1, k+1}$, $\bm{s}^{2, k+1}$ by solving Subproblem $0$, i.e., 
			\begin{equation}
	      \begin{split}
	      &(\bm{x}^{0, k+1}, \bm{s}^{1, k+1}, \bm{s}^{s, k+1})\\
	        &\quad  =\arg \min_{\bm{x}^0, \bm{s}^1, \bm{s}^2}  \mathbb{L}^0_{\rho} (\bm{x}^0, \bm{s}^1, \bm{s}^2, \bm{x}^{-0, k}, \bm{\lambda}^k, \bm{\gamma}^k, \bm{\eta}^k)  \\
	        & s. t.  \quad  \quad ~~\bm{x}^{0} \in \mathbb{R}.\\
	        &\quad \quad \quad  \quad  \bm{s}^1, \bm{s}^2 \geq \bm{0}.\\
	      \end{split}
    	\end{equation}
	\State For $i \in \mathcal{I}$, obtain $\bm{x}^{i, k+1}$ by solving Subproblem  $i$,  i.e., 
		\begin{eqnarray}
		\begin{split}
		&\bm{x}^{i, k+1}\!\!=\!\arg \min_{\bm{x}^i} \mathbb{L}^i_{\rho}\! (\bm{x}^i, \!\bm{x}^{-i, k}, \bm{s}^{1, k}, \bm{s}^{2, k},  \bm{\lambda}^k, \bm{\gamma}^k, \bm{\eta}^k)  \\
		& \quad s. t.  \quad  \quad \bm{x}^{i} \in \mathcal{X}^i\\
		\end{split}
		\end{eqnarray}		
		\State  Evaluate the residual of the coupled constraints $r(\bm{x}^{k+1},\bm{s}^{1, k+1}, \bm{s}^{2, k+1})$ according to (\ref{residual}). 
		If we have $r(\bm{x}^{k+1}, \bm{s}^{1, k+1}, \bm{s}^{2, k+1}) \leq \epsilon$, then stop and obtain the solution $\bm{x}^{k+1, *}$, othewise continue. 
		\State  Update the Lagrangian multipliers according to
		\begin{displaymath}
		\begin{split}
		&\bm{\lambda}^{i, k+1}\!=\!\!\bm{\lambda}^{i, k}\!+\!\rho\Big( \bm{A}_d^{ii}\bm{x}^{i, k+1} \!+\!\!\sum_{j\in \mathcal{N}_i} \bm{A}_d^{ij} \bm{x}^{j, k+1}\!\!-\!\!\bm{b}_d^i\Big), \\
		& \quad \quad \quad \quad \quad \quad \quad \quad \quad  \quad \quad \quad \quad \forall i\in \mathcal{I}. \\
		&\bm{\gamma}^{k+1}\!=\!\bm{\gamma}^k\!+\!\rho\Big( \sum_{i=0}^I \bm{B}_d^i \bm{x}^{i, k+1} \!+\!\bm{s}^{1, k+1}\Big)	\\
		&\bm{\eta}^{k+1}\!=\!\bm{\eta}^k\!+\!\rho\Big( \sum_{i=1}^I \bm{B}_d^i \bm{x}^{i, k+1}\!-\!\bm{c}_d \!+\!\bm{s}^{2, k+1}  \Big)
		\end{split}
		\end{displaymath}
		\State Set $k \rightarrow k+1$ and go to \textbf{Step} 3.
	\end{algorithmic}
\end{algorithm}

We use $\bm{x}^*= ( (\bm{x}^{0, *})^T,  (\bm{x}^{1, *})^T, \cdots, (\bm{x}^{I, *})^T)^T$ to represent the optimal solution of problem (\ref{P4}), which can be obtained through \textbf{Algorithm} \ref{ADAL} as introduced above.  
{{Whereas}}  we should note that  $\bm{x}^*$ may be infeasible to the original  problem (\ref{obj}).  Because  the recursive feasibility of the solution can't be guaranteed {{in the relaxed problem  (\ref{P4}).}}  To be specific,  the recursive equality $X^i_t=m^{zi}_t (T^i_t-T^c_t)$ ($\forall i \in \mathcal{I}, t \in \mathcal{T}$)  in (\ref{obj}) is relaxed in (\ref{P4}).  
Therefore, the remaining important  problem is how to  recover the recursive feasibility of the solution for  (\ref{obj}) based on the optimal solution of (\ref{P4})  while still guaranteeing  a satisfactory performance. 
To achieve this objective, an effective heuristic {{method}}  is developed by exploring  the special structures of  problem  (\ref{obj}).
Specifically, we note that  the decision variables  $X^i_t$  appear  in {{the}} global objective function of  (\ref{obj}), {{which ``dominate" the cost of the HVAC system compared with the decision variable $m^{zi}_t$ ($\kappa_f$ is a relatively small parameter)}}.  Meanwhile, the decision variables  $X^i_t$ determines the temperature of zone $i$ ($i \in \mathcal{I}$).  
{{Therefore,  when we try to recover the recursive feasibility of the solution,  if  high priority  is distributed to the decision variables $X^i_t$, the performance of the recovered solution (the cost of the HVAC system and the zone temperature) can be guaranteed. }} 
Following the above ideas, a heuristic method is proposed to recover the recursive feasibility of the solution, which is shown in \textbf{Algorithm} \ref{Heuristic1}. 
We use $\hat{m}^{zi}_t$, $\hat{T}^i_t$  and  $\hat{X}^i_t$ to represent the recovered solution for (\ref{obj}).  
We note that 
 a recovered control sequence ($\hat{m}^{zi}_t$, $\hat{T}^i_t$ and $\hat{X}^i_t$) for (\ref{obj})  is  obtained stage by stage. 
Specifically, the zone air flow rate $\hat{m}^{zi}_t$ is first determined by assuming  $\hat{X}^i_t=X^{i, *}_t$.
Meanwhile, the upper and lower bounds of  the zone air flow rate should be {{complied}}  (Step 5).
When the decision variable $\hat{m}^{zi}_t$ and $\hat{X}^i_t$ are determined,  the temperature of each zone should be  updated accordingly before we proceed to next stage (Step 7).  The recovered control sequence ($\hat{T}^i_t$, $m^{zi}_t$ and $X^i_t$) for each zone $i$ ($i \in \mathcal{I}$) can be obtained   by repeating this process until the end of the optimization horizon $\mathcal{T}$. 
  
\begin{algorithm}[h] 
	\caption{Heuristic Method to Recover the Recursive Feasibility of the Solution} \label{Heuristic1}
	\begin{algorithmic}[1]
		\State Obtain the optimal solution  $\bm{x}^*= ( (\bm{x}^{0, *})^T,  (\bm{x}^{1, *})^T, \cdots, $  $ (\bm{x}^{I, *})^T)^T$, 
		$\bm{X}^*= ( (\bm{X}^{0, *})^T,  (\bm{X}^{1, *})^T, \cdots, $  $ (\bm{X}^{I, *})^T)^T$ of the relaxed problem (\ref{P4}) according to \textbf{Algorithm} \ref{ADAL}.
	 \State Obtain the initial zone temperature $\hat{T}^i_0=T^i_0$ ($\forall i\in\mathcal{I}$). 
	\For{$t \in \mathcal{T}$}
	 \For{$i\in \mathcal{I}$}
	\State Determine the air flow rate of zone $i$ by
	\begin{equation}
		\begin{split}
		 &\overline{m}^{zi}_t=\min\big(\overline{m}^{zi}, X^{i, *}_t/(\hat{T}^i_t-T^c_t)\big), \\
		 & \hat{m}^{zi}_t=\max\big(\overline{m}^{zi}_t, \underline{m}^{zi})\big).
		\end{split}
	\end{equation}
	\State Determine  the auxiliary variable $\hat{X}^i_t$ of zone $i$ by
	\begin{equation}
	 \hat{X}^i_t=\hat{m}^{zi}_t (\hat{T}^i_t-T^c_t)
	\end{equation}
\State Update the temperature of zone  $i$ by
\begin{equation}
\hat{T}^{i}_{t+1}=A^{ii} \hat{T}^i_t+\sum_{j\in\mathcal{N}_i} A^{ij} \hat{T}^j_t +C^ii \hat{X}^i_t +D^{ii}_t
\end{equation}
\EndFor
\EndFor
\State Output the recovered solution $\hat{m}^{zi}_t$ and $\hat{T}^i_t$ ($\forall i \in \mathcal{I}, t\in \mathcal{T}$) of the original optimization problem (\ref{obj}). 
	\end{algorithmic}
\end{algorithm}

\section{Numeric  Results}

In this chapter, the performance of the decentralized approach  is evaluated through {{applications}}.  First of all, to evaluate and capture the suboptimality of the solution,  we compare the decentralized approach  with  a centralized method, in which an optimal solution for  a small-scale problem can be obtained.  
 Further,  the scalability of the decentralized approach  is validated through comparison with  the DTBSS method  \cite{radhakrishnan2016token}. 

\subsection{Performance Evaluation}
\begin{figure}
	\centering
	\includegraphics[width=3.5 in]{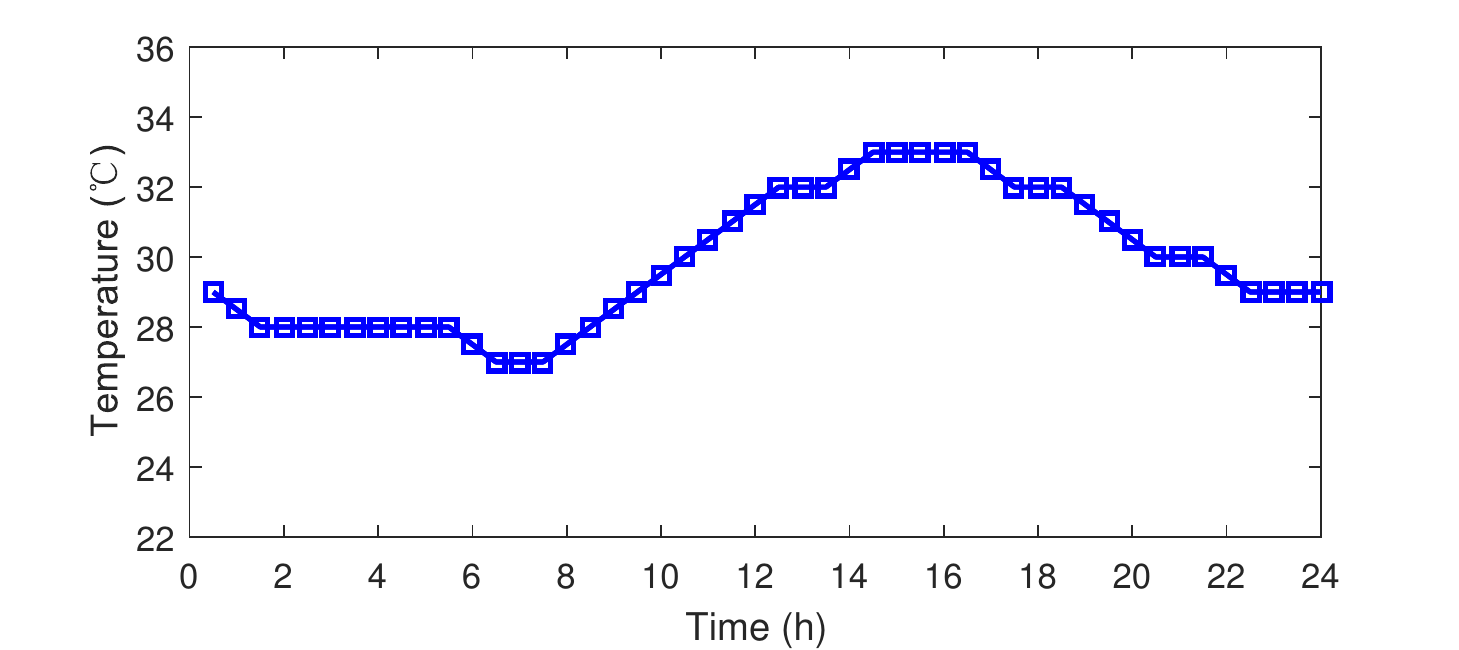}\\
	\caption{The outside air temperature} \label{Outside temperature}
\end{figure}

\begin{figure}
	\centering
	\includegraphics[width=3.5 in]{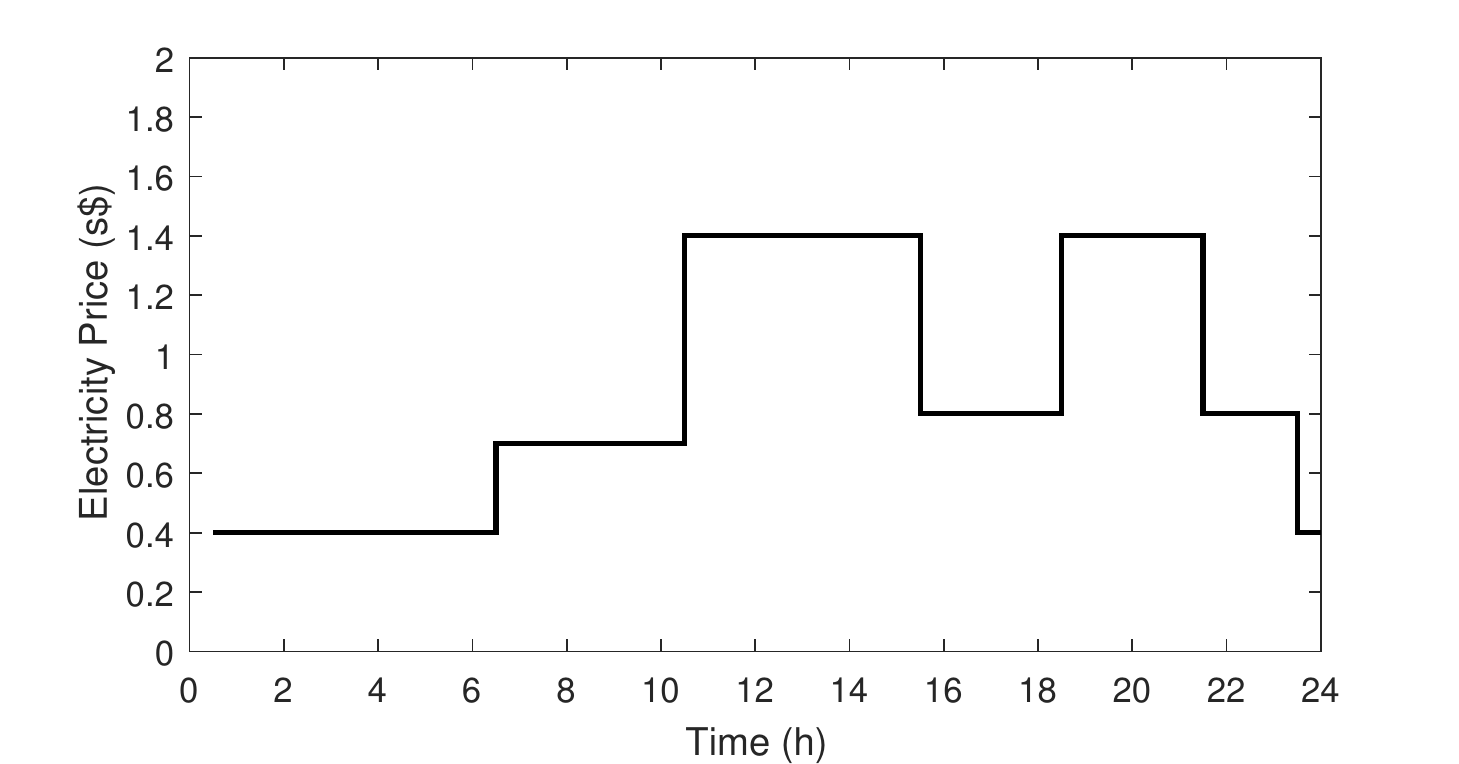}\\
	\caption{The TOU electricicty price}\label{Price}
\end{figure}

We first consider a small-scale case study with $I=2$ thermal zones.  Without loss of generality,   the comfortable temperature range  for all  zones  {{is}}  set as  $24$-$26^\circ$C  \cite{jia2017privacy, nagarathinam2015centralized} (we should note that the decentralized approach of this paper  can be applied to the case with various thermal comfort requirements for different zones).  
The outlet air temperature of the AHU is set as $T^c_t=15^\circ$C ($t \in \mathcal{T}$). 
Considering that the internal thermal load of each  individual zone is affected by various factors including the occupancy pattern, the usage of devices, etc.,  we randomly generate thermal load curves  for each  zone according to a  uniform distribution, i.e., $Q^i_t \sim U[0, 1] kW$ ($i\in \mathcal{I}, t\in \mathcal{I}$) {{in the following case studies. }}
In the two-zone case study,  there exist heat transfer between the two zones. 
And the initial zone temperature is  set as $[26, 28]^{\circ}$C, respectively. The outside air temperature usually fluctuates over the time as shown in Fig. \ref{Outside temperature}. 
The time-of-use (TOU) electricity price is shown in Fig.  \ref{Price}, which refers to \cite{xu2017pmv}. The other parameters are gathered in TABLE \ref{System  Parameters}. 
\begin{table}[htbp]
	\setlength{\abovecaptionskip}{-2pt}
	\setlength{\belowcaptionskip}{2pt}
	\scriptsize
	\centering
	\caption{System Parameters} \label{System  Parameters}
	\begin{tabular}{p{2cm}p{2cm}p{2cm}}
		\toprule[1.0pt]
		Param.  & Value   &Units\\
		\hline
		$C_i (i\in \mathcal{I})$   &  $1.375 \times 10^3$   & $J/(kg \cdot K)$\\
		$c_p$                                    & $1.012$                            & $kJ/(kg\cdot K)$\\    
		$R^{oi}$                                & $50$                                  & $kW/K$                \\
		$R^{ij} (i, j \in \mathcal{I})$           & $14$                    &$kW/K$                \\
		$k_f$                                         & $0.08$                            &-                              \\       
		$\eta$                                       & 1                                        &-                              \\
		\bottomrule[1.0pt]
	\end{tabular}
\end{table}

\begin{table}[htbp]
	\setlength{\abovecaptionskip}{-2pt}
	\setlength{\belowcaptionskip}{2pt}
	\scriptsize
	\centering
	\caption{Performance of the Centralized and Decentralized Methods} \label{two zone}
	\begin{tabular}{cccc}
		\toprule[1.0pt]
		\#Zones&Methods & Cost (s\$)   & Computation (s)\\
		\hline
		\multirow{2}{*}{$I=2$} &Centralized Method       &  24.01   & 52.71\\
		&Decentralized Approach                 &24.20                         &  3.09\\    
		\bottomrule[1.0pt]
	\end{tabular}
\end{table}

\begin{figure}
	\centering
	\includegraphics[width=3.5 in]{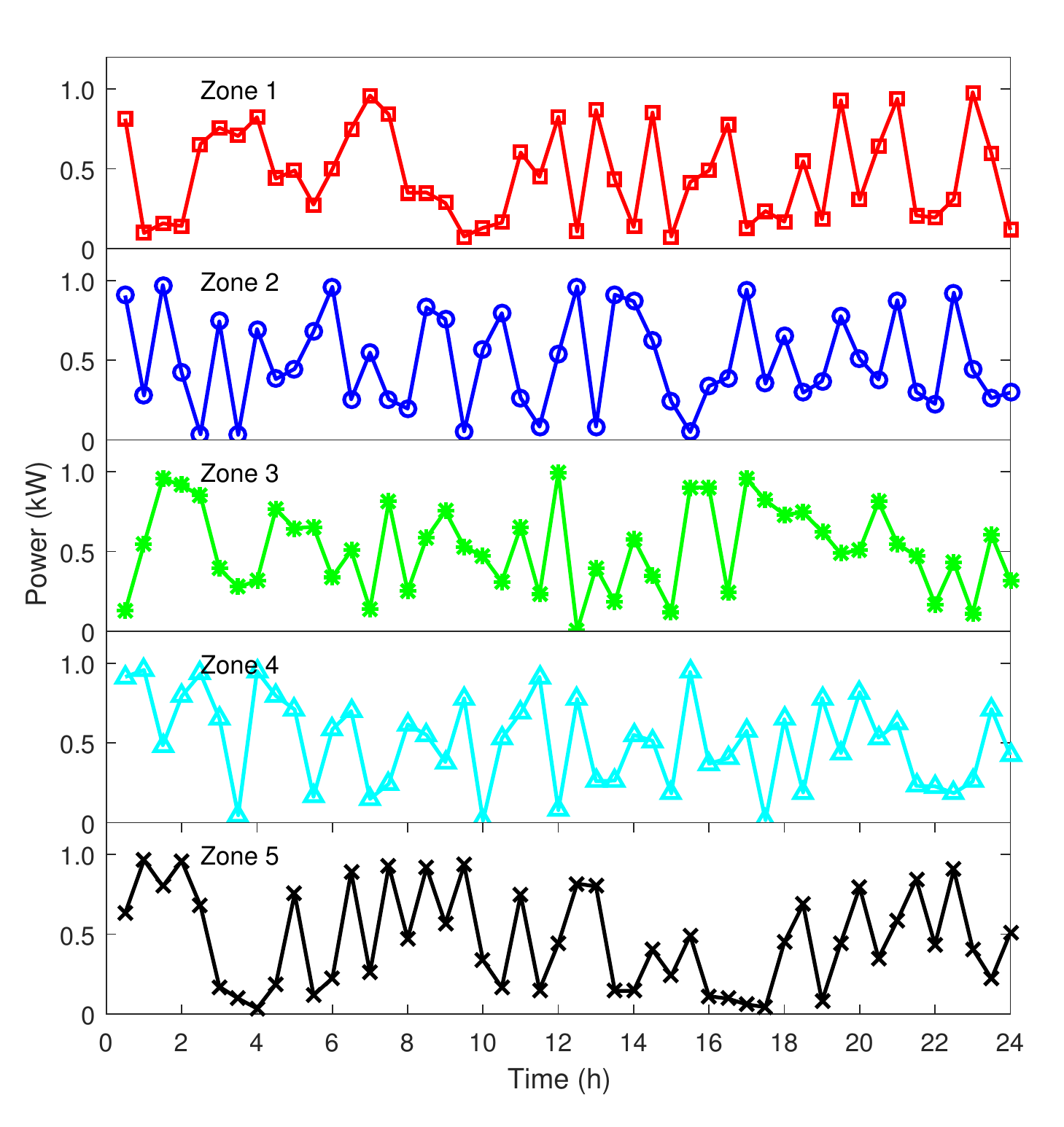}\\
	\caption{The thermal load curves for $I=5$ zones} \label{Thermal load}
\end{figure} 
\begin{table}[htbp]
	\setlength{\abovecaptionskip}{-2pt}
	\setlength{\belowcaptionskip}{2pt}
	\scriptsize
	\centering
	\caption{Performance Gap of the Decentralized Approach} \label{performance of different methods}
	\begin{tabular}{p{0.5 cm}p{3.5cm}p{1.2cm}}
		\toprule[1.0pt]
		\#Zones  & Method    &   Cost (s\$) \\
		\hline
		\multirow{3}{*}{$I=5$}               &Centralized Method   for (\ref{obj})                    &    60.20\\
		& Decentralized Approach for (\ref{obj}) &   62.05\\
		& ADAL for (\ref{P2}) &    60.63\\
		\bottomrule[1.0pt]
	\end{tabular}
\end{table}

In the two-zone case study, the performance of the 
decentralized approach  is compared with a centralized method,  in which  the optimal solution  of  problem  can be obtained  by solving the small-scale nonlinear  problem (\ref{obj}) using the IPOPT solver.
When the two methods are  applied, the total energy consumption cost of the HVAC system over the optimization horizon $\mathcal{T}$ and the average computation time for each  stage are contrasted in TABLE \ref{two zone}.    First, we note that the total energy consumption cost of the HVAC system under the two methods are comparable. Specifically, compared with the optimal solution (centralized method), there only exist a slight performance degradation  when the  decentralized approach  is adopted.  However, the average computation time for  each  stage is  apparently reduced.   {{Therefore,  we can preliminarily   conclude   that the decentralized approach  can  approach the optimal solution of the problem while reveals a substantial improvement on computation efficiency compared with a centralized method. }}

Further, to evaluate the performance of the decentralized approach,  we consider another case study with $I=5$ zones.  
In this case study, we assume there exist heat transfer  between any two of the zones.   Without loss of generality, the initial temperature for the $I=5$ zones are  set as $[26, 28, 28, 27, 27]^{\circ}$C, respectively.   Similarly,  the thermal load curves for the $I=5$ zones are randomly generated  according to the  uniform distribution,  which are shown in Fig. \ref{Thermal load}.  The other parameters can refer to the two-zone case study. 

\begin{figure}
	\centering
	\includegraphics[width=3.4 in]{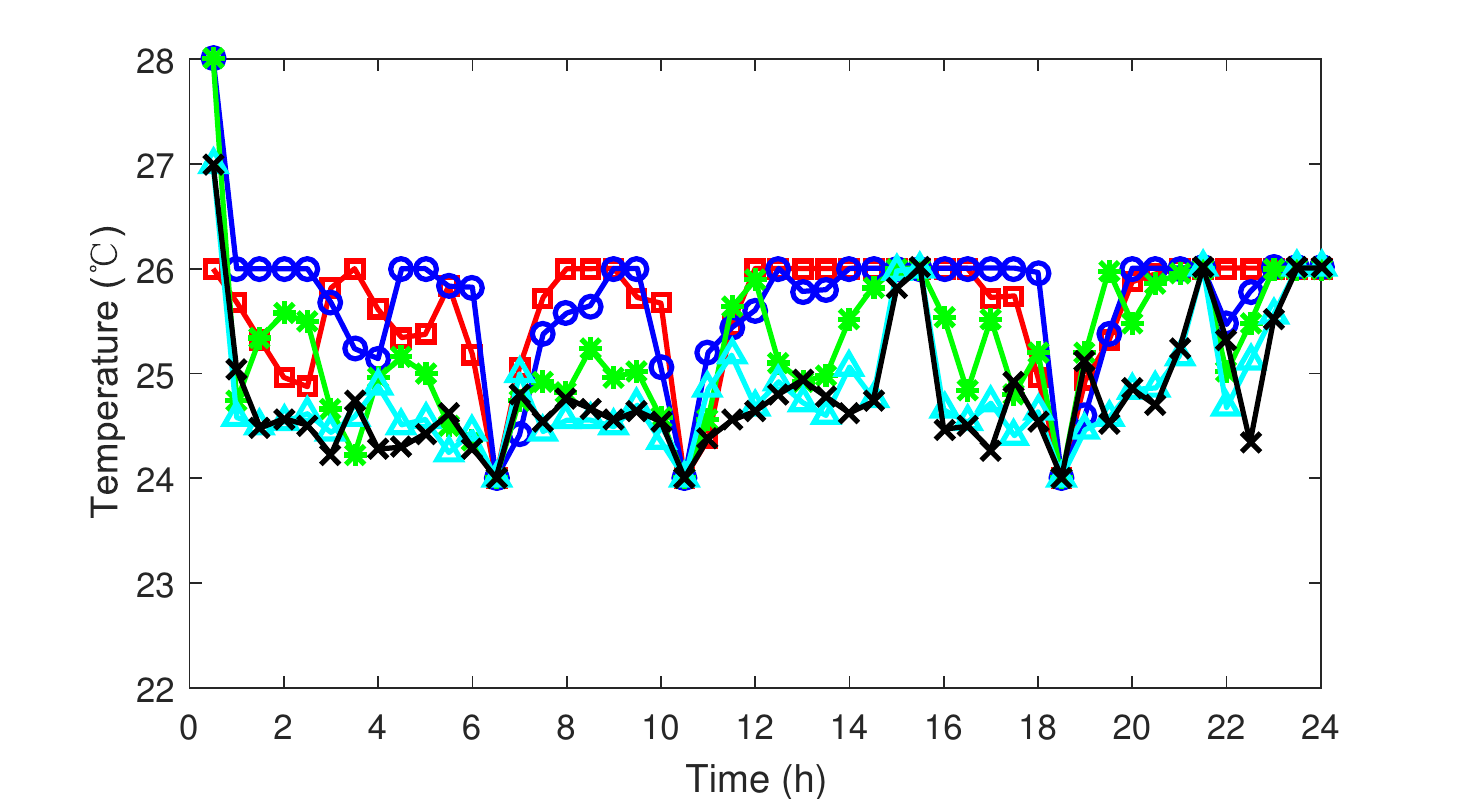}\\
	\caption{The zone temperature for $I$ zones} \label{Zone temperature}
\end{figure}

\begin{figure}
	\centering
	\includegraphics[width=3.4 in]{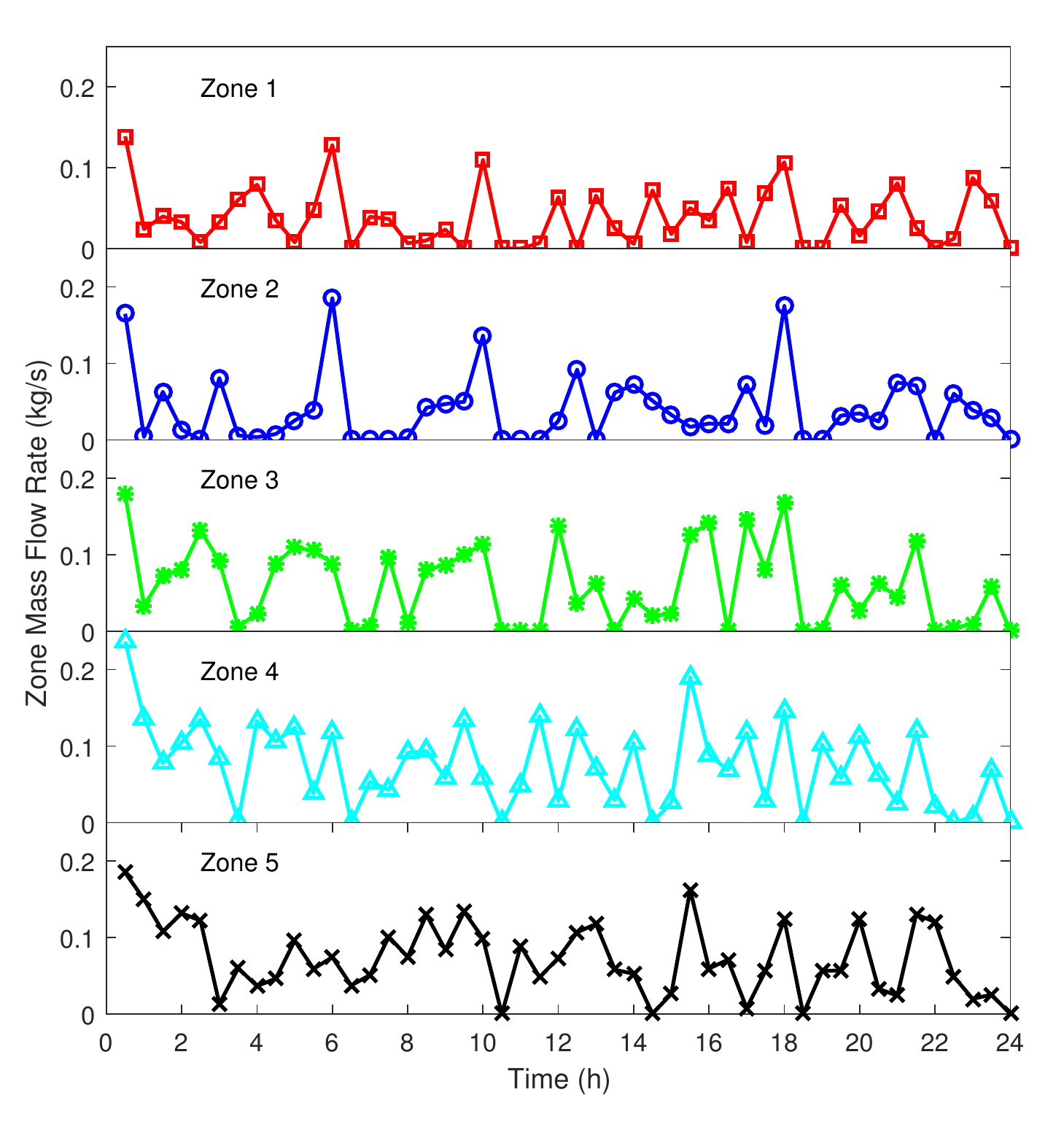}\\
	\caption{The air mass flow rate for $I$ zones} \label{Mass flow}
\end{figure}

 When the decentralized approach is applied, the curves of the zone  temperature are  plotted in Fig. \ref{Zone temperature}.  
 We see that over the optimization horizon, the  temperature of each individual zone can be  maintained  in the desired  comfortable temperature range 
 $24$-$26^{\circ}$C.   This implies that when the decentralized approach is employed, the {{thermal comfort}}  of each individual zone can be guaranteed.    Besides, from Fig. \ref{Zone temperature},   we  can observe some coincident valley points  regarding  the zone  temperature. 
And  these points correspond well to the  {{time instances}}  when the electricity price (as shown in Fig. \ref{Price}) {{begin to}}  rise.  This phenomenon is reasonable, because {{each zone}}  tends to  pre-cool the area   before the electricity price goes up to save the energy consumption cost. 
 Besides, the zone air flow rates for all  zones are plotted in Fig. \ref{Mass flow}.   We see that over the optimization horizon, the zone flow rates  for all zones are maintained in the  admissible  range  $[0, 0.5]~kg/s$.  
 To further evaluate the suboptimality of the solution, the energy consumption cost of the HVAC systems incurred by the decentralized approach and the centralized method are compared  in TABLE  \ref{performance of different methods}.  We note that the performance degradation is about $3.05\%$ when the decentralized approach is  applied. 
 Besides, the optimal cost of  the relaxed optimization  problem (\ref{P2}) is  also evaluated in the case study, which can serve  as the lower bound of the globally optimal cost for the original optimization problem (\ref{obj}).  We note that the optimal cost of the relaxed optimization problem (\ref{P2}) that  can be achieved by  the ADAL method (the second step of the decentralized approach)  is very close to  that of the centralized method. This implies that the performance loss of the decentralized approach is attributed to the heuristic method that adopted to recover the recursive feasibility of the solution (the third step of the decentralized approach).

\begin{figure}
	\centering
	\includegraphics[width=3.5 in]{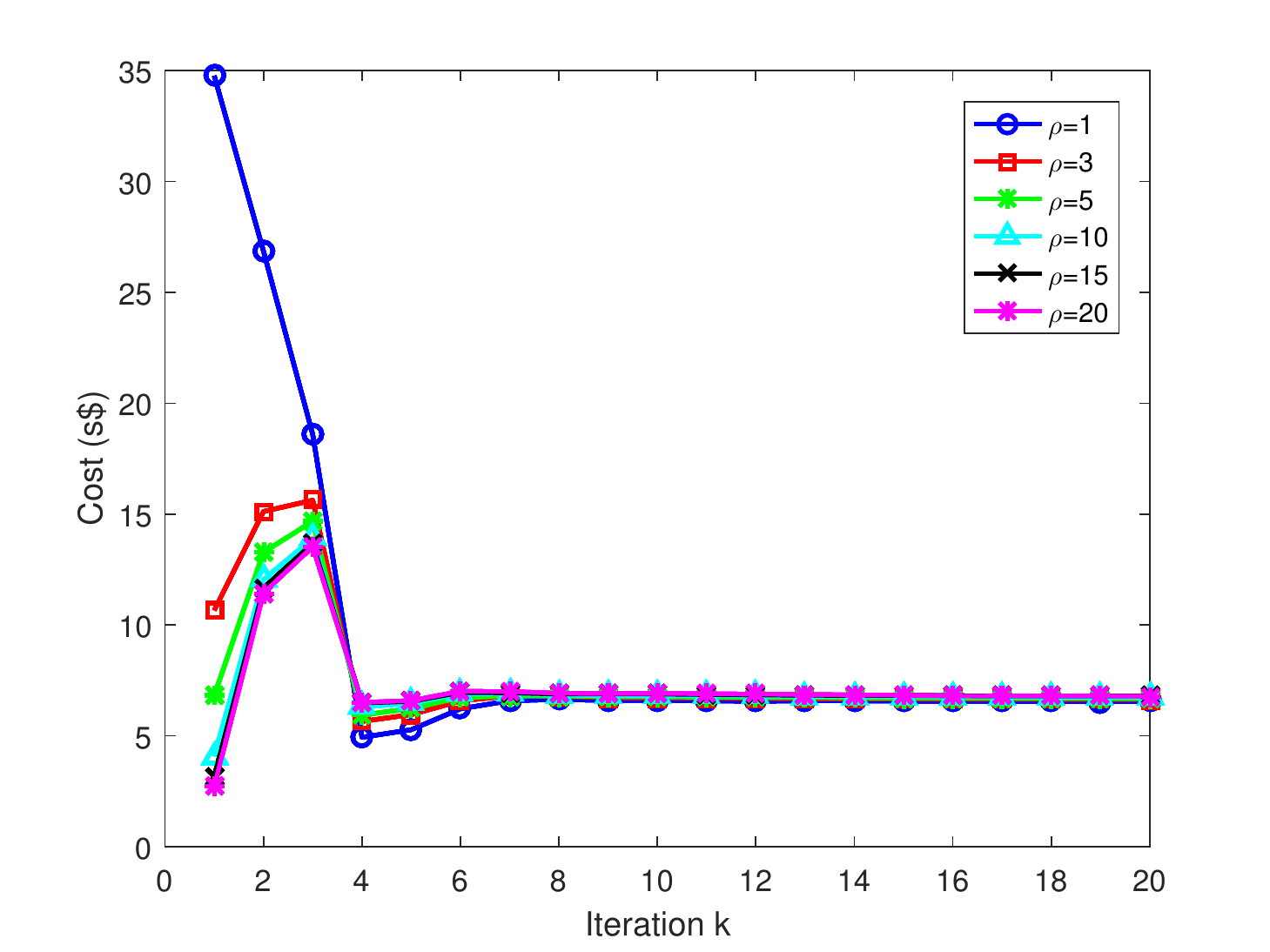}\\
	\caption{The convergence rate of objective function the decentralized approach  with different penalty parameters $\rho$. } \label{Convergence Rate}
\end{figure}

\begin{figure}
	\centering
	\includegraphics[width=3.5 in]{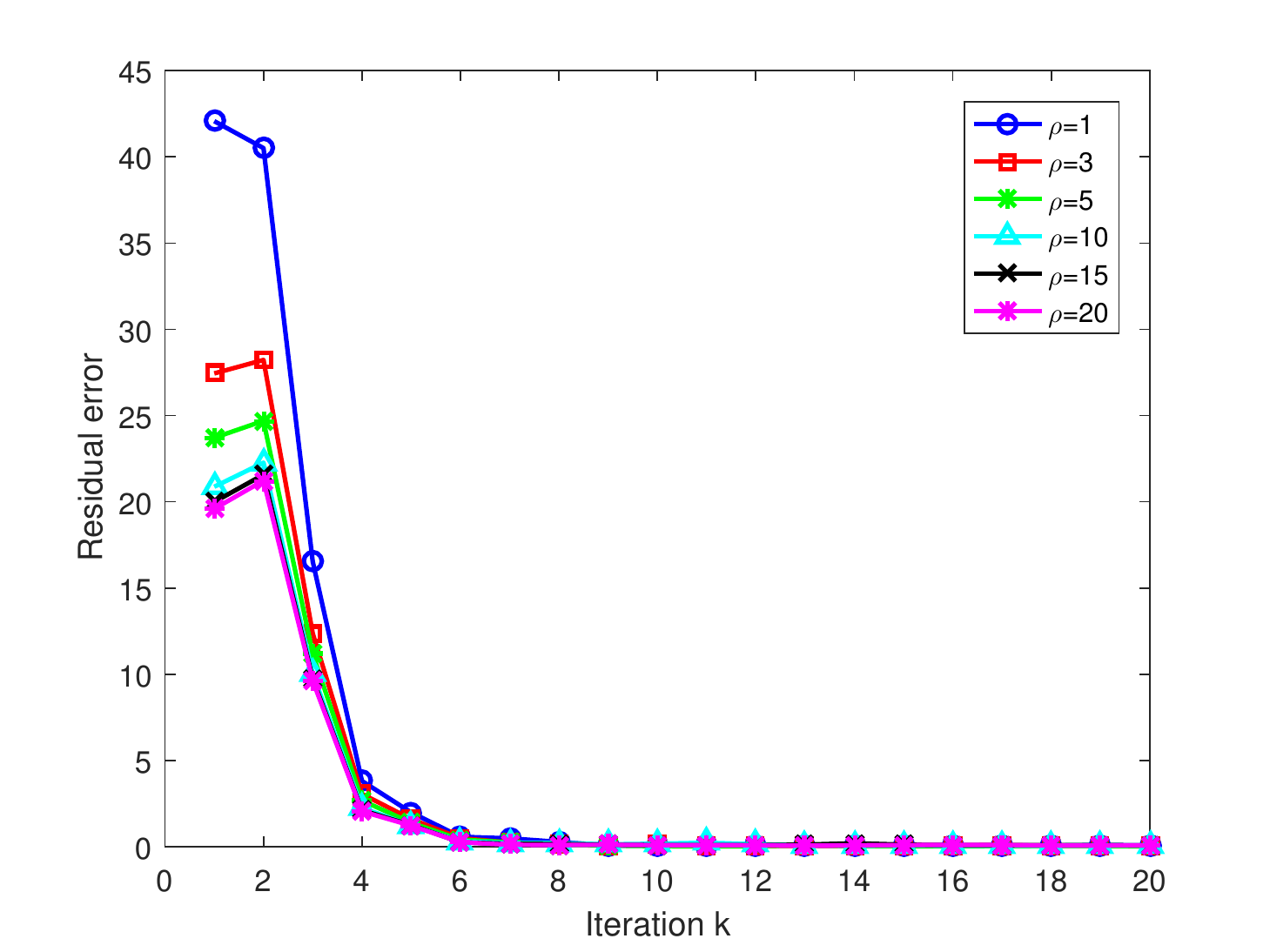}\\
	\caption{The convergence rate of the residual error of the decentralized approach  with different penalty parameters $\rho$. } \label{Residual error}
\end{figure}

Further,  to explore the possibility to accelerate the decentralized approach, we  analyze the convergence rate of the decentralized approach with different penalty  parameter $\rho$.    In the case study with $I=5$ zones, the convergence rates of the decentralized approach are compared under the  different penalty  parameters , i.e., $\rho=1$, $\rho=3$, $\rho=5$,  $\rho=10$,  $\rho=15$ and $\rho=20$.   The convergence rates {{regarding}}   the primal cost under the  different penalty parameters are contrasted  in Fig. \ref{Convergence Rate}.  We find that  a larger penalty parameter $\rho$ seems to  {{accelerate the  convergence rate  of the decentralized approach. }}
Besides, the convergence rate {{regarding}} the residual error of the coupled constraints are also compared   in Fig. \ref{Residual error}. 
{{Accordingly}}, we see that  the decentralized approach presents a  faster convergence rate  with a relatively larger $\rho$. 
{{Whereas}} when $\rho\geq 15$, the convergence rate of {{the decentralized approch both regarding}} the primal cost and the residual error are comparable. Therefore, to guarantee  a faster convergence rate, we set  $\rho=15$ in all  the following case studies.

\subsection{Scalability}

In this part, the scalability of the decentralized approach is evaluated through a series of  case studies with a larger number of zones. 
In particular, we compare the decentralized approach   with  the DTBSS method \cite{radhakrishnan2016token},  {{which}}   has been applied to the control of HVAC system in multi-zone buildings.  Generally speaking, the DTBSS method is a  heuristic hierarchical distributed method, in which the original  optimization problem (\ref{obj}) is divided into  three level subproblems.  Each of the subproblems consists of part of the overall cost function and a few related constraints.  To guarantee a fair comparison,  the  duct pressure constraints and the specification for the chiller efficiency mentioned in \cite{radhakrishnan2016token}  is relaxed in the case studies of this part.   {{Besides,  in the DTBSS method,  the thermal couplings among  the neighbouring zones are regarded as disturbances,  which are needed to be estimated  or measured ahead of the scheduling.   Therefore, to guarantee the implementation of the DTBSS method,  both  of the two decentralized methods  are carried out in a model predictive control (MPC) framework according to \cite{radhakrishnan2016token} in this part.  In this case, the disturbances  due to the thermal coupling   in the DTBSS method can be estimated based on the control trajectories computed over  the previous planning horizon. }}
 In general, when the DTBSS method is applied to  solve  problem (\ref{obj}),  the first step of the method  keeps unchanged with the objective to minimize the total energy consumption cost of the cooling power. And  this  subproblem  can be decomposed with respect to {{the zones.}}  Since the constraints due to duct pressure and chiller efficiency are omitted, the second step of the DTBSS method can be skipped.    And the third step is necessary with the objective to regulate the energy consumption cost of the fan within the AHU  as well as accommodating the coupling constraints related to the operation limit of the HVAC system.   That means  a nonlinear centralized optimization problem (the third subproblem) needs to be tacked  in the third step over each planning horizon.  The details of the DTBSS method can refer to \cite{radhakrishnan2016token}. 
In the two decentralized methods, the planning horizon is selected as $H=10$.  Since in the third step of the DTBSS method, a centralized problem is required to be tackled, the planning horizon of the third step of the DTBSS method is shorted to $H^{'}=5$ to reduce computation as suggested in  \cite{radhakrishnan2016token}.

We consider a number of case studies with different number of zones in this part, i.e., $I=5$,  $I=20$, $I=50$, $I=100$, $I=200$, $I=300$, $I=500$. 
We can use a network with $I+1$ nodes (Node $0$ represents the outside) to describe the thermal coupling between different  zones in the building. 
For any two node $i, j$ ($i, j \in \mathcal{I}$) of the network, we use $v^{ij}, v^{ji}=1$ to indicate that there exist thermal coupling (heat transfer) between zone $i$ and zone $j$, otherwise we have  $v^{ij}, v^{ji}=0$. Accordingly, we use $R^{ij}$ ($R^{ji}$)  to represent the thermal resistance between zone $i$ and zone $j$.
In  these case studies, we randomly generate some networks to describe the thermal connectivity of different  zones in the case studies. 
In accord with the common practice,  the maximum number of adjacent zones for each  zone is  set as  $4$.  {{The other paramters of the case studies in this part can refer to TABLE \ref{System  Parameters}.}}
When the two decentralized methods are applied, the 
energy consumption cost of the HVAC system  as well as the average  computation time for each individual zone at  each stage are contrasted  in TABLE  \ref{performance}. 

\begin{table}[htbp]
	\setlength{\abovecaptionskip}{-2pt}
	\setlength{\belowcaptionskip}{2pt}
	\scriptsize
	\centering
	\caption{Performance of the decentralized approach and the DTBSS} \label{performance}
	\begin{tabular}{c|rr|rr|r}
		\toprule[1.0pt]
		\multirow{2}{*}{\#Zones} & \multicolumn{2}{|c|}{DTBSS} & \multicolumn{2}{|c|}{ADAL}  &Cost~~~~~~\\
		\cline{2-5}
		&  Cost($s\$$)  & Time(s)  &  Cost($s\$$)  & Time(s) & Reduced (\%)\\
		\hline
		5               &                    63.57           &     2.11          &       62.05              &    1.39                     & 2.39       \\ 
		20            & 266.09              &         2.71                 &  258.79   &1.38                                   & 2.74     \\
		50              &                            713.07   &    4.47      &   687.23                    &  2.35                         &3.62           \\
		100                 &   $1.69\!\! \times\!\! 10^3$                            &      6.90   &     $1.61 \!\!\times\!\! 10^3$                 &     2.56                    & 4.73          \\
		200                &   $5.73\!\!\times\!\!10^3$  & 10.32    &  $5.10\!\!\times\!\!10^3$                   &   5.04                  & 11.00                                            \\
		300                 &  $1.51\!\!\times\!\! 10^4$ &17.89    &     $1.27\!\!\times\!\!10^4$            &   8.80           &     15.89                 \\
		500                &                    $5.57\!\!\times\!\!10^4$        &  28.01   &         $4.49\!\!\times10^4$                &       13.46                  &         19.39         \\
		\bottomrule[1.0pt]
	\end{tabular}
\end{table}

From TABLE  \ref{performance}, we see that  when the number of zones is relatively small (less than $20$),  the energy consumption costs  of the HVAC system under the  two  decentralized methods  are  comparable. There only exits a slight decrease ($2\%\!\!\sim\!\!3\%$) of the cost when the decentralized approach proposed in this paper is applied compared with the DTBSS method. However, with the increase of the number of  zones in the building,  the decentralized approach reveals  remarkably better performance   in reducing the total energy  cost.  Specifically, when the number of zones is $I=100$, the total energy consumption cost of the HVAC system can be reduced by about $11.0\%$ .  And when  the number of zones in the building  is increased to $I=300$,  the cost can be reduced by about o $15.89\%$. 
When the number of zones is $I=500$, the total energy consumption cost of the HVAC system  can be cut down by about $19.39\%$ compared with that of  the DTBSS method. 
The numeric results  imply  that the decentralized approach proposed in this paper outperforms the DTBSS method in reducing the total  cost of the HVAC system for buildings with a large number of zones.  {{In fact}},  it's not difficult to figure out the reasons for  the superior performance of the decentralized approach in reducing  the total  cost of the  HVAC system compared with the DTBSS method.
Generally,  in the DTBSS method, the total energy consumption cost of the HVAC system (the cooling power and the fan power)  is distributed in  the subproblems of different level. {{More Specifically, }} ths DTBSS method first distributes high priority to minimize the energy consumption cost {{of  the cooling  power related to each zone}}  without considering the energy consumption cost cause by  the fan within the AHU.
After that the energy consumption of the fan is regulated in the third step provided that the cooling demand of each individual zone in Step one  is ensured.  
We should note that the third step of the DTBSS method may result in an increase of the cost for  cooling power by reducing the cost of  fan power. 
However, in the decentralized approach {{proposed in this paper}},  both the energy consumption  cost caused by  the cooling power and  the fan power is  coordinated  at the same time.  Therefore, the decentralized approach tends to reduce more  energy consumption cost for the HVAC system.  
Besides,  {{as aforementioned}}, we find that  when the number of zones is relatively small (less than $20$),  there doesn't exist any evident difference between the cost under the two decentralized methods.  The reason is attributed to the fact that when the number of zones is small,  the fan power within  the AHU has a much smaller effect on the total  cost of the HVAC system ( the fan power depends on the cube of the total zone air flow rate) compared with the cooling power, which results in the small performance gap {{between}}  the two methods.  However, for buildings with  a large number of zones,  the effect of the fan power on the total cost of the HVAC system will be increased rapidly. Therefore,  it will make an apparent difference in the total cost  of the HVAC system when the energy consumption of the cooling  power and the fan power  is  coordinated in the decentralized approach.   

Further,   we compare the computation time of the two decentralized methods.  Considering that both of the two decentralized methods can be carried out in a parallel mode,   the average computation time for each individual zone at each  stage is   contrasted  in TABLE \ref{performance}.     {{We find that the two decentralized methods are both  computationally efficient. }}
However,  the decentralized approach outperforms the DTBSS method with less average computation time required by each individual zone at each stage.   The reasons are attributed to several aspects: 1) the decentralized approach presents a fast convergence due to the  fast convergence of the ADAL method, which can be seen in Fig. \ref{Convergence Rate}. 2) in the decentralized approach, each subproblem corresponding to each individual zone is a samll-scale QP problem, which can be  efficiently tacked by many existing toolbox.  However, in the first step of the DTBSS method,  a nonlinear  and nonconvex subproblem is needed to be tacked by each individual zone.  3) in the DTBSS method,  a centralized problem related to  all the thermal zones  is needed to be solved in the third step with the objective   to regulate the fan power of the AHU and the coupled constraints  of the problem. 
{{Therefore,}}  with the number of zones increased, the computation related to the DTBSS method  will increase rapidly.  
Thus, the numeric results imply that the decentralized approach proposed in this paper demonstrates a  satisfactory performance both  in reducing  the cost of the HVAC system along with improving  the computational efficiency.

\section{Conclusion}
This paper studies the control of the HVAC systems  in multi-zone buildings with the objective to reduce the energy consumption cost  while guaranteeing  the zone thermal comfort.    Considering that centralized methods   are usually time-consuming or  intractable for buildings with a large number of zones,  an efficient decentralized approach based on the Accelerated Distributed Augmented Lagrangian (ADAL) method \cite{chatzipanagiotis2017convergence} is developed in this paper.  
Through comparison with a centralized method,  we find that the decentralized approach can  approach the optimal solution of the problem.  Besides, to evaluate the performance and the scalability of the decentralized approach, we compare it  with the Distributed Token-Based Scheduling Strategy (DTBSS) method  \cite{radhakrishnan2016token}, which has been developed for the control of HVAC systems for multi-zone buildings.  
We find that when the number of zones is relatively small (less than $20$),  there only exist a narrow performance gap ($2\%\!\!\sim\!\!3\%$) regarding the cost of the HVAC system.  
{{However, with an increasing number of zones, }}  the decentralized approach can reduce a considerable amount of the energy consumption cost for the HVAC system compared with the DTBSS method. In particular, when the number of zones is $I=500$, the total energy consumption cost of the HVAC system can be  reduced by about $19.39\%$. Moreover, we find that the decentralized  method shows better performance with an apparantly decrese of the computation time.


\appendices
\section{Problem P3}
The temperature dynamics of zone $i$ can be described as 
\begin{equation} \label{dynamics}
\begin{split}
&T^i_{t+1}  
= \left(            
\begin{array}{ccc}   
A^{ii} & 0  &C^{ii} \\  
\end{array}
\right)          
\left(            
\begin{array}{c}   
T^i_{t} \\  
m^{zi}_{t} \\  
X^i_{t}
\end{array}
\right) \\
&\quad \quad \quad \quad \quad +\sum_{j \in \mathcal{N}_i}
\left(            
\begin{array}{ccc}   
A^{ij} & 0  &0 \\  
\end{array}
\right)          
\left(            
\begin{array}{c}   
T^j_{t} \\  
m^{zj}_{t} \\  
X^j_{t}
\end{array}
\right) 
+
D^{ii}_t\\
\end{split}
\end{equation}
If we define  $\overline{A}^{ii}=  \left(            
\begin{array}{ccc}   
A^{ii} & 0  &C^{ii} \\  
\end{array}
\right)          
$,  $\overline{A}^{ij}=   \left(            
\begin{array}{ccc}   
A^{ij} & 0  &0 \\  
\end{array}
\right)$
and $\overline{D}^{ii}_t\!\!=\!\!-D^{ii}_t$, and the decision variable $\bm{x}^i=( (\bm{x}^i_0)^T, (\bm{x}^i_1)^T, \cdots (\bm{x}^i_{T-1})^T )^T$ ($\forall i \in \mathcal{I}$) with $\bm{x}^i_t=(T^i_t, m^{zi}_t, X^i_t)$,  the temperature dynamics for all the thermal zones can be combined by 
\begin{equation}\label{(30)}
\begin{split}
& \left(      
\begin{array}{ccccc}   
\overline{A}_{ii} & -\bm{I}_1  & \cdots & \cdots & \cdots\\  
\bm{0} &  \overline{A}_{ii} & -\bm{I}_1 & \cdots &\cdots\\  
\bm{0} & \bm{0} & \overline{A}_{ii} & -\bm{I}_1  & \cdots\\
\end{array}
\right) 
\bm{x}^i\\
&\quad \quad  +\sum_{j\in \mathcal{N}_i}        
\left(      
\begin{array}{cccccc}   
\overline{A}_{ij} & \bm{0}  & \cdots & \cdots & \cdots\\  
\bm{0} &  \overline{A}_{ij} & \bm{0} & \cdots & \cdots \\  
\bm{0} & \bm{0} & \overline{A}_{ij} & \bm{0}  & \cdots\\
\end{array}
\right)\bm{x}^j \\
&\quad \quad \quad \quad +  \left(      
\begin{array}{c}   
\overline{D}^{ii}_0\\  
\vdots\\
\overline{D}^{ii}_{T-1}\\ 
\end{array}
\right)=\bm{0}
\end{split}
\end{equation}

Further, we define\\
$$\bm{A}_d^{ii}= \left(      
\begin{array}{cccccc}   
\overline{A}^{ii} & -\bm{I}_1  & \bm{0} & \cdots & \cdots& \cdots\\  
\bm{0} &  \overline{A}^{ii} & -\bm{I}_1 & \bm{0} &\cdots & \cdots \\  
\bm{0} & \bm{0} & \overline{A}^{ii} & -\bm{I}_1  & \bm{0} & \cdots \\
\end{array}
\right) \in \mathbb{R}^{(T-1) \times 3T}, $$
$\bm{A}_d^{ij}=\left( \begin{array}{ccccc}   
\overline{A}^{ij} & \bm{0}  & \cdots & \cdots & \cdots\\  
\bm{0} &  \overline{A}^{ij} & \bm{0} & \cdots & \cdots\\  
\bm{0} & \bm{0} & \overline{A}^{ij} & \bm{0} & \cdots \\
\end{array}
\right) \in \mathbb{R}^{(T-1) \times  3T}, \quad \quad 
$ and 
$\bm{b}_d^i=(\overline{D}^{ii}_0, ~\overline{D}^{ii}_1, \cdots,  \overline{D}^{T-1}_{ii} )^T\in \mathbb{R}^{(T-1)}$

Thus, the dynamics in (\ref{(30)}) is eqivalent to  
\begin{equation}
\begin{split}
\bm{A}_d^i \bm{x}^i +\sum_{j\in \mathcal{N}_i} \bm{A}_d^{ij} \bm{x}^j=\bm{b}_d^i
\end{split}
\end{equation}

Similarly, the coupled constraints in (\ref{14h}) can be written as
\begin{equation} \label{(32)}
\begin{split}
\sum_{i=0}^{I}
\left(
\begin{array}{ccc} 
0 & 1 & 0
\end{array}
\right)
\left(
\begin{array}{c} 
T^i_t\\
m^{zi}_t\\
X^i_t
\end{array}
\right) - Y_t \leq 0
\end{split}
\end{equation}

If we define $\overline{B}^i=(0~~1~~0)$ ($\forall i\in\mathcal{I}$)  and $\overline{B}^0=(-1)$. The coupled constraints in (\ref{(32)}) over the optimization horizon  $\mathcal{T}$ can be collected as 
\begin{equation} \label{(33)}
\begin{split} \sum_{i=0}^I 
\left(
\begin{array}{cccc}
\overline{B}^i & \bm{0} &\bm{0}   & \cdots\\
\bm{0} & \overline{B}^i & \bm{0} &\cdots \\
\bm{0}  &\bm{0} & \overline{B}^i   &\cdots\\
\end{array}
\right) \bm{x}^i \leq  \bm{0}
\end{split}
\end{equation}

If we define $\bm{B}_d^i\!\!=\!\!\left(\begin{array}{cccc}
\overline{B}^i & \bm{0} &\bm{0} & \cdots \\
\bm{0} & \overline{B}^i & \bm{0}  & \cdots\\
\bm{0}  &\bm{0} & \overline{B}^i  & \cdots  \\
\end{array} \right) \in \mathbb{R}^{T \times 3T}$ ($\forall i\in \mathcal{I}$), 
(\ref{(33)}) is equivalent to 
\begin{equation}
\begin{split}
\sum_{i=0}^I \bm{B}_d^i \bm{x}^i \leq 0
\end{split}
\end{equation}

Accordingly, the constraints  (\ref{14i}) can be described as 
\begin{equation}
\begin{split}
\sum_{i=1}^I \bm{B}_d^i \bm{x}^i \leq \bm{c}_d
\end{split}
\end{equation}
where we have 
$\bm{c}_d =(\overline{m}, \overline{m} \cdots \overline{m})^T \in \mathbb{R}^T$

\ifCLASSOPTIONcaptionsoff
  \newpage
\fi

\bibliographystyle{ieeetr}
\bibliography{reference}
\end{document}